\documentclass[preprint,prd,aps,showpacs,showkeys,nofootinbib]{revtex4}
\usepackage{graphicx}
\usepackage{dcolumn}
\usepackage{bm}
\usepackage{amsmath} % 必须加载的数学宏包
\usepackage{color}
\usepackage[dvipsnames]{xcolor}
\usepackage{amssymb}
\usepackage{float}
\usepackage{subfigure}
\usepackage{tabularx}
\usepackage{booktabs} % 用于专业表格线
\definecolor{black-blue}{RGB}{77,116,175}
\definecolor{black-yellow}{RGB}{231,162,33}
\definecolor{black-green}{RGB}{144,180,58}
\definecolor{black-red}{RGB}{246,95,50}
\begin{document}
\title{Analyzing the magnitude of two loop corrections to the muon magnetic dipole moment in the mass insertion approximation}
\author{Shu-Min Zhao$^{1,2,3}$\footnote{zhaosm@mail.nankai.edu.cn}, Song Gao$^{1,2,3}$,
Xing-Xing Dong$^{1,2,3,4}$\footnote{dongxx@hbu.edu.cn}, Ming-Yue Liu$^{1,2,3}$}

\affiliation{$^1$ Department of Physics, Hebei University, Baoding 071002, China}
\affiliation{$^2$ Hebei Key Laboratory of High-precision Computation and Application of Quantum Field Theory, Baoding, 071002, China}
\affiliation{$^3$ Hebei Research Center of the Basic Discipline for Computational Physics, Baoding, 071002, China}
\affiliation{$^4$ Departamento de Fisica and CFTP, Instituto Superior T$\acute{e}$cnico, Universidade de Lisboa,
Av.Rovisco Pais 1,1049-001 Lisboa, Portugal}

\date{\today}

\begin{abstract}
With the development of muon magnetic dipole moment (MDM) experiments, particularly the high-precision measurements at Fermilab National Accelerator Laboratory (FNAL), experimental data have become increasingly precise.
Up to now, the deviation of muon MDM between the experimental data and the standard model prediction is still existing,
which may be caused by the new physics contribution.
The two loop supersymmetric diagrams can produce important corrections to muon MDM.
For many two loop diagrams, we analyse the order of the contribution to select the important two loop
diagrams and neglect the tiny one, which is in favor of distinguishing the important two loop diagrams from all the two loop diagrams.
 In our analysis, we use the mass insertion approximation. There is no rotation matrix during the study and the resulting factors become more intuitive, so the mass insertion method is more suitable for this study of ours.
\end{abstract}

\keywords{two loop, mass insertion approximation, muon MDM}

\maketitle

\section{introduction}
 Muon magnetic dipole moment(MDM) has attracted the interests of physicists for many years.
 For muon MDM, the SM contributions\cite{SMmuonMDM,SMMDM,SMmuonMDM1,SMmuonMDM2} can be divided into several parts:
QED contribution is dominant\cite{SMmuonMDM,QEDEW}; Electroweak and hadronic contributions are also important\cite{QEDEW, HVP}.
 The latest experiment data for muon MDM is reported by the Fermilab National Accelerator Laboratory (FNAL)\cite{FA}.
Combined with the previous data, the world average value of the departure between experiment
data and the SM prediction is \cite{cao}
\begin{eqnarray}
\Delta a_\mu=a^{exp}_\mu-a^{SM}_\mu=38(63)\times 10^{-11}.
 \end{eqnarray}
One explanation is that the deviation $\Delta a_\mu$ is caused by the new physics.
A computation using SUSY for such small deviation is still interesting.
Through the symmetry of fermions and bosons, SUSY can cancel out the divergences with each other, making the theoretical calculations more accurate, which provides a natural solution to the hierarchy problem in the Higgs mass corrections.
By the running of the renormalization group equation, the three gauge coupling constants intersect at one point at
the very high energy scale around $10^{16}$ GeV. The lightest neutralino can be a dark matter candidate.
SUSY has new particles beyond the SM, and these new particles can bring
 new corrections to muon g-2.
Even the deviation for the muon MDM between the experimental data and the SM prediction is minimal, the SUSY may provide a more self-consistent explanation by introducing new particles, such as supersymmetry partner particles.

Up to now, SM has achieved
great successes with the detection of 125 GeV Higgs boson\cite{Higgs125}. However, SM can not provide cold dark matter candidate,
which is one shortcoming of SM.
Therefore, physicists extend the SM to obtain new models.
The minimal supersymmetric extension of the standard model(MSSM)\cite{MSSM} is one of the most popular models, which has been studied
for many years. Based on the MSSM, there are also many extended models\cite{newmodel,U1X} with
added gauge group and superfields especially right-handed neutrino and Higgs superfields.
In these new models, there are many particles contributing to muon MDM at two loop level\cite{twoloopSUSY1,twoloopSUSY2}.
Therefore, so many two loop triangle diagrams for $\mu\rightarrow\mu\gamma$ must be faced.
It is very important to distinguish the order of the each type two loop corrections for muon MDM\cite{Zhao2020}.

The two loop diagrams in the SM have been well studied to obtain the electroweak corrections to muon MDM\cite{SMmuonMDM}.
In the new physics models such as 2HDM\cite{THDtwoloop} and MSSM,
some types of two loop diagrams are researched carefully by the authors\cite{twoloopSUSY1,twoloopSUSY2}.
For the two loop diagrams, the Barr-Zee type\cite{feng08prd}, the rainbow type\cite{feng08npb},
the diamond type\cite{feng06prd,Zhao2022jhep} are all of interest and are calculated analytically.
Furthermore, the decoupled results are also shown apparently with the supposition that all
supersymmetric particles are heavy and of the same mass.
When the internal fermion or sfermion are very heavy, the two loop diagrams with fermion/sfermion
loops have large factors\cite{SSL}. The exact results for
the photonic supersymmetric two loop corrections can be found in Ref.\cite{twoloopphoton}.

In our previous work, we have researched the factors of the two loop contributions
to muon MDM in mass eigenstate\cite{Zhao2020}.
It benefits the selection of the two loop diagrams.
However, the rotation matrixes of the mass mixing matrixes are not considered in the analysis,
which may influence the factor obviously.
For the one loop contributions from $(\chi^0, \tilde{L})$ and $(\chi^\pm, \tilde{\nu})$ in the mass eigenstate,
 there is an enhancement factor $\frac{m_{\chi}}{m_\mu}$.
 However, this is not the truth, because the combined rotation matrixes suppress the results.
 Considering the effects of these factors as a whole, one can obtain the
 factor $\frac{m_\mu^2}{M_{NP}^2}\tan\beta$ in the end, which is analyzed by the authors in detail\cite{MEtoMIA}.
 It is supposed that the SUSY particles are all heavy and with the same mass $M_{NP}$.
 The detailed analysis for the one loop contributions is shown here.

 In the MSSM, we present the general results of the one loop diagrams in the following form \cite{PRDimportant,MEtoMIA}
\begin{eqnarray}
&&a_{\mu}^{\mathrm{MSSM},1\mathrm{L}} = a_{\mu}^{\chi^{0}} + a_{\mu}^{\chi^{\pm}},\nonumber\\
&&a_{\mu}^{\chi^{0}} = \frac{1}{48\pi^{2}} \sum_{i,m}\frac{m^2_{\mu}}{m_{\tilde{\mu}_{m}}^{2}}
\Big\{ - \frac{|n_{i m}^{L}|^{2} + |n_{i m}^{R}|^{2}}{4} F_{1}^{N}(x_{i m}) + \frac{m_{\chi_{i}^{0}}}{m_{\mu}} \mathrm{Re}[n_{i m}^{L} n_{i m}^{R}] F_{2}^{N}(x_{i m}) \Big\},\nonumber\\&&
a_{\mu}^{\chi^{\pm}} = \frac{1}{48\pi^{2}} \sum_{k} \frac{m^2_{\mu}}{m_{\tilde{\nu}_{\mu}}^{2}}
\Big\{ \frac{ |c_{k}^{L}|^{2} + |c_{k}^{R}|^{2} }{4} F_{1}^{C}(x_{k})
 + 2\frac{m_{\chi_{k}^{\pm}}}{m_{\mu}} \mathrm{Re}[c_{k}^{L} c_{k}^{R}] F_{2}^{C}(x_{k}) \Big\}.\label{oneloop}
\end{eqnarray}
 $i = 1\ldots4$, $k = 1,2$ and $m = 1,2$ respectively denote the neutralino, chargino and smuon indices.
  The mass ratios are defined as $x_{i m} = m_{\chi_{i}^{0}}^{2} / m_{\tilde{\mu}_{m}}^{2}$, $x_{k} = m_{\chi_{k}^{\pm}}^{2} / m_{\tilde{\nu}_{\mu}}^{2}$.  The concrete forms of the couplings are
\begin{eqnarray}
&&n_{i m}^{L} = \frac{1}{\sqrt{2}} (g_{1} N_{i 1} + g_{2} N_{i 2}) U_{m 1}^{\tilde{\mu}*} - y_{\mu} N_{i 3} U_{m 2}^{\tilde{\mu}*},
\nonumber\\&&
n_{i m}^{R} = \sqrt{2} g_{1} N_{i 1} U_{m 2}^{\tilde{\mu}} + y_{\mu} N_{i 3} U_{m 1}^{\tilde{\mu}},
\nonumber\\&&
c_{k}^{L} = - g_{2} Z^{1k}_+,~~~~~~~~
c_{k}^{R} = y_{\mu}Z^{2k}_-.
\end{eqnarray}

The smuon mass squared  matrix is diagonalized by $U^{\tilde{\mu}}$.
While, $N_{i 1}$ is the rotation matrix to diagonalize the neutralino mass matrix.
To obtain the mass eigenvalues of chargino, two rotation matrixes $Z_-$ and $Z_+$ are needed, whose concrete
 forms are in the Eqs.(\ref{chimass})$\cdots$ (\ref{ZZ}).

The one loop functions in Eq.(\ref{oneloop}) are shown as
\begin{eqnarray}
&&F_{1}^{N}(x) = \frac{2}{(1 - x)^{4}} [1 - 6x + 3x^{2} + 2x^{3} - 6x^{2} \log x],\nonumber\\&&
F_{2}^{N}(x) = \frac{3}{(1 - x)^{3}} [1 - x^{2} + 2x \log x],
\nonumber\\&&
F_{1}^{C}(x) = \frac{2}{(1 - x)^{4}} [2 + 3x - 6x^{2} + x^{3} + 6x \log x],
\nonumber\\&&
F_{2}^{C}(x) = \frac{3}{(1 - x)^{3}} [-3 + 4x - x^{2} - 2 \log x].
\end{eqnarray}
These functions are normalized as $F_{i}^{j}(1) = 1$.

In Eq.(\ref{oneloop}), on the surface
there is the factor $\frac{m_{\chi^\pm}}{m_\mu}~(\frac{m_{\chi^0}}{m_\mu})$
in the second term for $a_\mu^{\chi^\pm}$( $a_\mu^{\chi^0}$ ).
In fact, the terms linear in $m_{\chi^{0,\pm}}$ are not
enhanced by a factor $m_{\chi^{0,\pm}} / m_{\mu}$ compared to the other terms.
Let's take $\frac{m_{\chi^\pm}}{m_\mu}$ as an example.
From Eq.(\ref{oneloop}), we discuss the total effects of the following terms
\begin{eqnarray}
&&\sum_{k} \frac{m^2_{\mu}}{m_{\tilde{\nu}_{\mu}}^{2}}
 \frac{m_{\chi_{k}^{\pm}}}{m_{\mu}} \mathrm{Re}[c_{k}^{L} c_{k}^{R}] F_{2}^{C}(x_{k})
\nonumber\\&& =\sum_{k} \frac{m^2_{\mu}}{m_{\tilde{\nu}_{\mu}}^{2}}
 \frac{m_{\chi_{k}^{\pm}}}{m_{\mu}}
 \mathrm{Re}[- g_{2} V_{k 1}y_{\mu} U_{k 2}] F_{2}^{C}(x_{k})\nonumber\\&&
 \propto \frac{m_{\mu}^2} {M_{NP}^2} \tan \beta.\label{OLFX}
\end{eqnarray}
Here, $y_{\mu}$ is the muon Yukawa coupling.
When we take the concrete forms of the parameters in Eq.(\ref{OLFX}),
the factor $\frac{m_{\mu}^2}{  M_{NP}^2} \tan \beta$ is obtained.
That is to say the terms linear in $m_{\chi^{0,\pm}}$ are enhanced merely by a factor $\tan \beta$ from the muon Yukawa coupling.

The two loop SUSY contributions also have similar features as the one loop condition\cite{MEtoMIA}.
 Considering the combined rotation matrixes, the factor $\frac{m_{\chi}}{m_\mu}$ on the surface in the two loop corrections
 is suppressed in the total analysis.
 To get the factors more clearly and thoroughly,
 we use the mass insertion approximation(MIA)\cite{MIA,MIA1}
 to analyse the two loop triangle diagrams
 contributing to muon MDM.
 With MIA method we use the electroweak interaction eigenstate and treat perturbatively the mass insertions.
 In the MIA approach it is not necessary to include rotation mass matrices.
 Therefore, the results are clear for the vertex couplings in MIA method.
 The energy scale of the new physics is supposed as $\Lambda_{NP}$, and we take  $\Lambda_{NP}\sim M_{NP}$.

 In the section II, the used suppositions and analysis method are shown in detail.
 The detailed example and obtained factors for the two loop diagrams are collected in section III.
 The last section is devoted to the discussion and conclusion.

\section{the method}

There are many SUSY models, where MSSM is very representative.
Though the analysis is  employed in the MSSM, it is also adaptable to other SUSY models and even non-SUSY models.
Instead of the mass eigenstate, we use MIA,
which is calculated in the electroweak interaction eigenstate.
This method has advantages in the study of lepton flavor violation to show important parameters apparently\cite{MIA1,DXX2024}.
Similarly, the advantages can also be reflected in the muon MDM analysis\cite{MIA}.
There is not the rotation matrixes,
and the obtained factors become more intuitive.

There are a great many two loop SUSY triangle diagrams for $\mu\rightarrow \mu \gamma$.
To save space in the text, we plot the two loop self-energy diagrams, from
 which the two loop triangle diagrams can be obtained by attaching a photon on the
 internal charged lines in all possible ways. The one loop diagrams in electroweak state are
  much more than the one loop diagrams in mass eigenstate.
  Similarly, the two loop self-energy diagrams using MIA are
 much more than the two loop self-energy diagrams
 in the mass eigenstate.
 Therefore, we show some examples,
 and the other two loop diagrams are still denoted in the mass eigensate.

\begin{figure}[h]
\setlength{\unitlength}{1mm}
\centering
\includegraphics[width=6in]{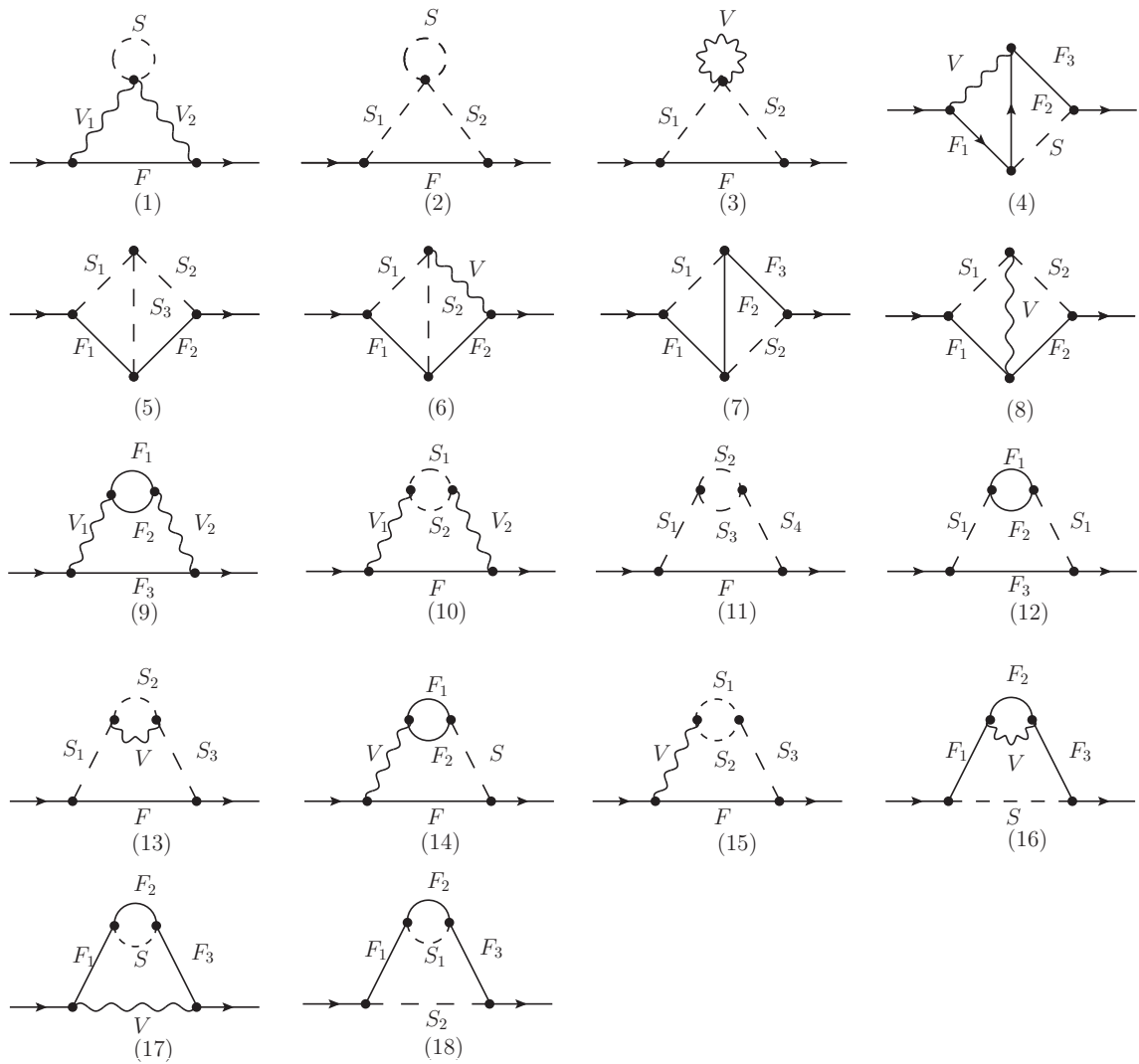},
\vspace{0cm}
\caption[]{The two loop self-energy diagrams for muon MDM in mass eigenstate.}\label{TLSEME}
\end{figure}

In mass eigenstate, the two loop self-energy diagrams are plotted in the Fig.\ref{TLSEME}.
To simplify the discussion,
we use the notations that $S, V$ and $F$ represent virtual scalar boson, vector boson and Dirac(Majorana) particle respectively.
 Some diagrams(Fig.\ref{TLSEME}(4), Fig.\ref{TLSEME}(6), Fig.\ref{TLSEME}(14), Fig.\ref{TLSEME}(15))
 have their Hermitian diagrams to produce the same contribution to muon MDM, and we do not show them here.
  Their Hermitian diagrams can be obtained by exchanging the vector boson and the corresponding scalar boson.

In previous works, the authors show the decoupling results for dominant one-loop SUSY contributions to muon MDM,
which all have the typical factor $\frac{m_\mu^2}{M_{NP}^2}\tan\beta$ with all SUSY particles masses being $M_{NP}$\cite{MEtoMIA,MIA}.
To simplify the discussion,
we also use the supposition that all SUSY particles have the same mass $M_{NP}$ in this work.
Generally speaking, the masses of neutralino and sneutrino should be around 500 GeV, or even heavier.
The slepton mass is no less than 700 GeV\cite{PDG2024}. The lightest chargino mass is at the scale of 1100 GeV.
The very heavy SUSY particles are colored particles including scalar quark and gluino,
whose masses are not smaller than 1500 GeV.  For the sake of simplicity,
we suppose that the SUSY particles approximately have the same mass $M_{NP}\sim 1000 ~{\rm GeV}$.
From a numerical point of view, $1000 ~{\rm GeV}$ is about the central value for the most SUSY particles' masses from 500 GeV
to 1500 GeV, which is a reasonable approximation. This assumption greatly reduces the complexity of the analysis.

\begin{figure}[h]
\setlength{\unitlength}{1mm}
\centering
\includegraphics[width=2.5in]{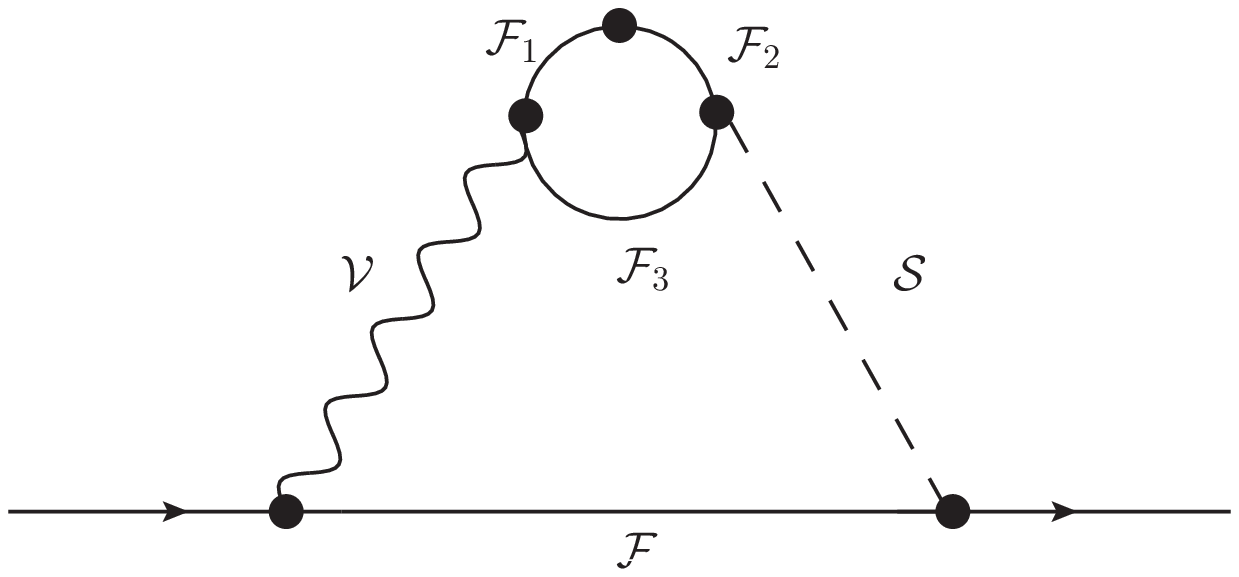},
\vspace{0cm}
\caption[]{The two loop self-energy diagram in electroweak eigenstate corresponding to Fig.\ref{TLSEME}(14).}\label{fmtMIA1}
\end{figure}

The sum of all the two loop triangle diagrams($\mu\rightarrow\mu\gamma$)
corresponding to each two loop self-energy diagram($\mu\rightarrow\mu$) satisfies the Ward-identity\cite{ward,ward1}.
We take two examples to explain the conversion between
the two loop self-energy diagrams in mass eigenstate and the corresponding diagrams using MIA in electroweak eigenstate.
For the Fig.\ref{TLSEME}(14), we take $V=\gamma, ~F=\mu, ~S=h^0,~ F_1=F_2=\chi^\pm$,
whose corresponding diagram processed by MIA is shown in the Fig.\ref{fmtMIA1}
with $\mathcal{V}=\gamma, ~\mathcal{F}=\mu, ~\mathcal{S}=h^0_d,~ \mathcal{F}_1=\tilde{W}^\pm,~ \mathcal{F}_2=\tilde{H}^\pm,~ \mathcal{F}_3=\tilde{W}^\pm$.
This is the Barr-Zee type two loop diagram with heavy fermion sub-loop,
which contributes to muon MDM with the typical factor $\frac{m_\mu^2}{M^2_{NP}}\tan\beta$\cite{MEtoMIA}.
Large $\tan\beta$ can improve this type two loop diagram contribution.

\begin{figure}[h]
\setlength{\unitlength}{1mm}
\centering
\includegraphics[width=2.5in]{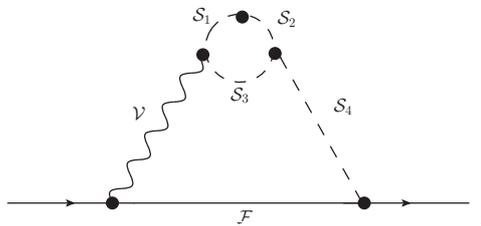},
\vspace{0cm}
\caption[]{The two loop self-energy diagram in electroweak eigenstate corresponding to Fig.\ref{TLSEME}(15).}\label{fmtMIA2}
\end{figure}

The Barr-Zee type two loop diagram with scalar sub-loop
in MIA is shown in Fig.\ref{fmtMIA2}. The corresponding particles are respectively
$\mathcal{V}=\gamma, ~\mathcal{F}=\mu, ~\mathcal{S}_1=\tilde{t}_R,
~ \mathcal{S}_2=\tilde{t}_L,~ \mathcal{S}_3=\tilde{t}_R, ~\mathcal{S}_4=h^0_d$.
As discussed in Ref.\cite{MEtoMIA}, this diagram has the typical parameter $\frac{m_\mu^2}{M_{NP}^2}\tan\beta\frac{m_t^2}{m_W^2}$,
which is analyzed in the next section.

\section{the analysis}

In order to obtain representative analysis, we study the two loop diagrams in the MSSM\cite{MSSM}.
Inserting MSSM particles into the two loop self-energy diagrams in generic form,
a lot of two loop diagrams are produced from each type generic diagram.
To save space in the text, we use the two loop diagrams in mass eigenstate
to represent two loop diagrams in electroweak state.
Each two loop diagram in mass eigenstate
corresponds to several two loop diagrams in electroweak state, which are analysed
by MIA method in fact. Then, the results are obtained by MIA method, and
the two loop diagrams are not so many to plot.

\subsection{The techniques for the two loop function}
Here, we show the techniques for the two loop diagram calculation.
The used formulas to simplify the Feynman integrals are\cite{Zhao2022jhep,Feng2009npb}
{\small
 \begin{eqnarray}
&&\int\!\!\!\frac{d^Dk_1d^Dk_2}{(2\pi)^{2D}}\frac{(k_{1\mu}k_{1\nu},~k_{1\mu}k_{2\nu})}{\mathcal{D}_0}
\rightarrow \int\!\!\!\frac{d^Dk_1d^Dk_2}{(2\pi)^{2D}} \frac{(k^2,~k_1\cdot k_2)g_{\mu\nu}}{D\mathcal{D}_0},
%%%%%%%%%%%%%%%%%%%%%%%%%%%%%%%%%%%%%%%%%%%%%
\nonumber\\
&&\int\!\!\!\frac{d^Dk_1d^Dk_2}{(2\pi)^{2D}}\frac{(k_{1\mu}k_{1\nu}k_{1\rho}k_{1\sigma},~k_{1\mu}k_{1\nu}k_{1\rho}k_{2\sigma})}{\mathcal{D}_0}
\rightarrow\int\!\!\!\frac{d^Dk_1d^Dk_2}{(2\pi)^{2D}}\frac{(k_1^4,~k_1^2k_1\cdot k_2)T_{\mu\nu\rho\sigma}}{D(D+2)\mathcal{D}_0},\nonumber\\
&&\int\!\!\!\frac{d^Dk_1d^Dk_2}{(2\pi)^{2D}}\frac{k_{1\mu}k_{1\nu}k_{2\rho}k_{2\sigma}
}{\mathcal{D}_0}\nonumber\\
 &&\rightarrow\int\frac{d^Dk_1d^Dk_2}{(2\pi)^{2D}}\frac{1}{\mathcal{D}_0}
 \Big(\frac{D(k_1\cdot
k_2)^2-k_1^2k_2^2}{D(D-1)(D+2)}T_{\mu\nu\rho\sigma}-\frac{(k_1\cdot
k_2)^2-k_1^2k_2^2}{D(D-1)}g_{\mu\nu}g_{\rho\sigma}\Big),
%%%%%%%%%%%%%%%%%%%%%%%%%%%%%%%%%%%%%%%%%%%%%%%%%%%%
\nonumber\\
&&\int\!\!\!\frac{d^Dk_1d^Dk_2}{(2\pi)^{2D}}\frac{k_{1\mu}k_{1\nu}k_{1\rho}k_{1\sigma}
k_{2\alpha}k_{2\beta}}{\mathcal{D}_0}\nonumber\\
 &&\rightarrow\int\!\!\!\frac{d^Dk_1d^Dk_2}{(2\pi)^{2D}}\frac{1}{\mathcal{D}_0}
 \Big(\frac{D(k_1\cdot
k_2)^2k_1^2-k_1^4k_2^2}{D(D+2)(D+4)(D-1)}T_{\alpha\beta\mu\nu\rho\sigma}-\frac{(k_1\cdot
k_2)^2k_1^2-k_1^4k_2^2}{D(D+2)(D-1)}g_{\alpha\beta}T_{\mu\nu\rho\sigma}\Big),\nonumber\\
&&\int\frac{d^Dk_1d^Dk_2}{(2\pi)^{2D}}\frac{k_{1\mu}k_{1\nu}k_{1\rho}k_{2\sigma}
k_{2\alpha}k_{2\beta}}{\mathcal{D}_0}\nonumber\\
 &&\rightarrow\int\!\!\!\frac{d^Dk_1d^Dk_2}{(2\pi)^{2D}}\frac{1}{\mathcal{D}_0}
 \Big(\frac{(D+1)k_1^2k_2^2(k_1\cdot
k_2)-2(k_1\cdot
k_2)^3}{D(D+2)(D+4)(D-1)}T_{\alpha\beta\mu\nu\rho\sigma}-\frac{k_1^2k_2^2(k_1\cdot
k_2)-(k_1\cdot k_2)^3}{D(D+2)(D-1)}\nonumber\\
&&\times[g_{\mu\sigma}(g_{\nu\alpha}
g_{\rho\beta}+g_{\nu\beta}g_{\rho\alpha})+g_{\mu\alpha}(g_{\nu\sigma}g_{\rho\beta}
+g_{\nu\beta}g_{\rho\sigma})+g_{\mu\beta}(g_{\nu\sigma}g_{\rho\alpha}
+g_{\nu\alpha}g_{\rho\sigma})]\Big)\label{tihuan},
 \end{eqnarray}}
with the time-space dimension $D=4-2\epsilon$. $T_{\mu\nu\rho\sigma}$, $T_{\mu\nu\rho\sigma\alpha\beta}$ and
$\mathcal{D}_0$ are shown as
\begin{eqnarray}
&&T_{\mu\nu\rho\sigma}=g_{\mu\nu}g_{\rho\sigma}+g_{\mu\rho}g_{\nu\sigma}
+g_{\mu\sigma}g_{\nu\rho},\nonumber\\&&
T_{\mu\nu\rho\sigma\alpha\beta}=g_{\mu\nu}T_{\rho\sigma\alpha\beta}+g_{\mu\rho}T_{\nu\sigma\alpha\beta}
+g_{\mu\sigma}T_{\nu\rho\alpha\beta}+g_{\mu\alpha}T_{\nu\rho\sigma\beta}+g_{\mu\beta}T_{\nu\rho\sigma\alpha},\nonumber\\&&
\mathcal{D}_0=(k_1^2-m_1^2)(k_2^2-m_2^2)\Big((k_1-k_2)^2-m_0^2\Big).
\end{eqnarray}

The odd rank terms of $k_1$ and $k_2$ in the
loop momenta are abandoned, because their integrals are zero.
With the following decomposition formula,
we reduce the complicated two-loop integrals to the simple  two-loop vacuum
integrals with one $1/\mathcal{D}_0$ and one-loop integrals
\begin{eqnarray}
&&\frac{1}{(k^2-m_1^2)}\frac{1}{(k^2-m_2^2)}=\frac{1}{m_1^2-m_2^2}\left(
\frac{1}{k^2-m_1^2}-\frac{1}{k^2-m_2^2}\right)\label{chai},\nonumber
\\&&
\frac{1}{(k^2-m^2)^n}=\frac{1}{(n-1)!}\left(\frac{\partial}{\partial
m^2}\right)^{n-1}\frac{1}{k^2-m^2}\label{dao}.
\end{eqnarray}

The two-loop vacuum integral is obtained\cite{Feng2009npb}
\begin{eqnarray}
&&\Lambda_{{\rm RE}}^{4\epsilon}\int\int{d^Dk_1d^Dk_2\over(2\pi)^{2D}}{1\over
(k_1^2-m_1^2)(k_2^2-m_2^2)((k_1-k_2)^2-m_0^2)}
\nonumber\\&&=
{\Lambda^2\over2(4\pi)^4}{\Gamma^2(1+\epsilon)\over(1-\epsilon)^2}
\Big({4\pi x_{R}}\Big)^{2\epsilon}
\Big\{-{1\over\epsilon^2}\Big(x_0+x_1+x_2\Big)
+{1\over\epsilon}\Big(-(x_0+x_1+x_2)\nonumber\\&&+2(x_0\ln x_0+x_1\ln x_1+x_2\ln x_2)\Big)
+2(x_0\ln x_0+x_1\ln x_1+x_2\ln x_2)
\nonumber\\&&-2(x_0+x_1+x_2)
-x_0\ln^2x_0-x_1\ln^2x_1-x_2\ln^2x_2-\Phi(x_0,x_1,x_2)\Big\},
\label{vacuum}
\end{eqnarray}
with
\begin{eqnarray}
&&\Phi(x,y,z)=(x+y-z)\ln x\ln y+(x-y+z)\ln x\ln z
\nonumber\\
&&\hspace{1.7cm}+(y+z-x)\ln y\ln z+{\rm sign}(\lambda^2)\sqrt{|\lambda^2|}\Psi(x,y,z)\;,
\label{phi}
\\&&
\lambda^2(x,y,z) = x^2 + y^2 + z^2 - 2xy - 2yz - 2xz.\label{Lambda}
\end{eqnarray}
The concrete form of $\Psi(x,y,z)$ reads as:

$1.~~\lambda^2>0,\;\sqrt{y}+\sqrt{z}<\sqrt{x}$:
\begin{eqnarray}
&&\Psi(x,y,z)=2\ln\Big({x+y-z-\lambda\over2x}\Big)
\ln\Big({x-y+z-\lambda\over2x}\Big)-\ln{y\over x}\ln{z\over x}
\nonumber\\
&&\hspace{2.2cm}
-2L_{i_2}\Big({x+y-z-\lambda\over2x}\Big)
-2L_{i_2}\Big({x-y+z-\lambda\over2x}\Big)+{\pi^2\over3}\;.
\label{aeq2}
\end{eqnarray}
 $L_{i_2}(x)$ is the spence function;

 $2.~~\lambda^2>0,\;\sqrt{x}+\sqrt{z}<\sqrt{y}$:
\begin{eqnarray}
&&\Psi(x,y,z)={\rm Eq.}(\ref{aeq2})(x\leftrightarrow y)\;;
\label{aeq3}
\end{eqnarray}

$3.~~\lambda^2>0,\;\sqrt{x}+\sqrt{y}<\sqrt{z}$:
\begin{eqnarray}
&&\Psi(x,y,z)={\rm Eq.}(\ref{aeq2})(x\leftrightarrow z)\;;
\label{aeq4}
\end{eqnarray}

 $4.~~\lambda^2<0$:
\begin{eqnarray}
&&\Psi(x,y,z)=2\Big\{Cl_2\Big(2\arccos(
{-x+y+z\over2\sqrt{yz}})\Big)
+Cl_2\Big(2\arccos({x-y+z\over2\sqrt{xz}})\Big)
\nonumber\\
&&\hspace{2.2cm}
+Cl_2\Big(2\arccos({x+y-z\over2\sqrt{xy}})\Big)\Big\}\;.
\label{aeq5}
\end{eqnarray}
$Cl_2(x)$ denotes the Clausen function.

We show the used partial functions of $\Phi(x_1,x_2,x_3)$.

1. The function $\frac{\partial \Phi}{\partial x_1}(x_1,x_2,x_3)$
\begin{eqnarray}
&&\frac{\partial\Phi}{\partial x_1}(x_1,x_2,x_3)=\frac{x_1-x_2-x_3}{\sqrt{\lvert\lambda^2\rvert}}
\mathrm{\Psi}(x_1,x_2,x_3)+\log(x_1)\log(x_2)\nonumber\\
&&\hspace{3.0cm}+\log(x_1)\log(x_3)+2\log(x_1)-\log(x_2)\log (x_3).
\end{eqnarray}

2. The function $\frac{\partial^2\varPhi}{\partial x_1^2}(x_1,x_2,x_3)$
\begin{eqnarray}
&&\frac{\partial^2\varPhi}{\partial x_1^2}(x_1,x_2,x_3)=-\frac{4 x_2 x_3 }{(\lvert\lambda^2\rvert)^{3/2}}\mathrm{\Psi}(x_1,x_2,x_3)
-\frac{2 (-x_1+x_2+x_3) \log (x_1)}{\lambda^2}\nonumber\\
&&
~~ ~~ ~~ ~~ ~~ ~~ ~~ ~~ ~~ +\frac{\log (x_2)+\log (x_3)+2}{x_1}
-\frac{ (x_3-x_1-x_2) (x_3-x_1+x_2)\log (x_2)}{x_1 \lambda^2}\nonumber\\
&&
~~ ~~ ~~ ~~ ~~ ~~ ~~ ~~ ~~
+\frac{ (x_1-x_2+x_3) (-x_1+x_2+x_3)\log (x_3)}{x_1 \lambda^2}.
\end{eqnarray}

3. The function $\frac{\partial^2\Phi}{\partial x_1\partial x_2}(x_1,x_2,x_3)$
\begin{eqnarray}
&&\frac{\partial^2\varPhi}{\partial x_1\partial x_2}(x_1,x_2,x_3)
=\frac{1}{\lambda^2}\Big(-2(x_1 - x_2 + x_3)\log(x_1)- 2(x_3-x_1 + x_2)\log(x_2) \nonumber\\&&
\hspace{3.3cm}+ 4x_3 \log(x_3)\Big)
+ \frac{2x_3(x_1 + x_2 - x_3)}{\lvert \lambda^2 \rvert^{3/2}} \mathrm{\Psi}(x_1, x_2, x_3).
\end{eqnarray}

4. The function $\frac{\partial^3\Phi}{\partial x_1^3}(x_1,x_2,x_3)$
\begin{eqnarray}
&&\frac{\partial^3\varPhi}{\partial x_1^3}(x_1,x_2,x_3)
=\frac{12 x_2 x_3 (x_1 - x_2 - x_3)}{\lvert \lambda^2 \rvert^{5/2}}\mathrm{\Psi}(x_1, x_2, x_3)- \left(
  \frac{24 x_2 x_3}{\lambda^4}
  + \frac{2}{\lambda^2}\right)\log(x_1)
 \nonumber\\&&~~ ~~ ~~ ~~ ~~ ~~ ~~ ~~ ~~ ~~  + \left(
  \frac{(x_3 - x_1)^2 - x_2^2}{x_1^2 \lambda^2}
  + \frac{2 x_2}{x_1 \lambda^2}
  - \frac{12 x_2 x_3 (-x_1 - x_2 + x_3)}{x_1 \lambda^4}
\right)\log(x_2)
\nonumber\\&&~~ ~~ ~~ ~~ ~~ ~~ ~~ ~~ ~~ ~~
+ \left(
  \frac{2 x_3 (2 x_1 + x_2 - x_3)}{x_1^2 \lambda^2}
  + \frac{12 x_2 x_3 (x_1 - x_2 + x_3)}{x_1 \lambda^4}
  + \frac{1}{x_1^2}
\right)\log(x_3)\nonumber\\&&~~ ~~ ~~ ~~ ~~ ~~ ~~ ~~ ~~ ~~
- \frac{2 [ (x_3 - x_2)^2 - x_1 (x_2 + x_3) ]}{x_1^2 \lambda^2}
- \frac{\log(x_2) + \log(x_3)}{x_1^2}.
\end{eqnarray}

5. The function $\frac{\partial^3\varPhi}{\partial x_1^2 \partial x_2}(x_1,x_2,x_3)$
\begin{eqnarray}
&&\frac{\partial^3\varPhi}{\partial x_1^2 \partial x_2}(x_1,x_2,x_3)=\frac{2(x_2 - x_1 - x_3)}{x_1 \lambda^2}
+ \left( \frac{2}{\lambda^2} + \frac{12 x_3(x_1 + x_2 - x_3)}{\lambda^4} \right) \log(x_1)
\nonumber\\&&~~ ~~ ~~ ~~ ~~ ~~ ~~ ~~ ~~ ~~ ~~  - \left( \frac{24 x_2 x_3}{\lambda^4} + \frac{2(x_1 + x_3)}{x_1 \lambda^2} \right) \log(x_2)
\nonumber\\&&~~ ~~ ~~ ~~ ~~ ~~ ~~ ~~ ~~ ~~ ~~  + \left( \frac{8 x_3(x_2 - x_1 + x_3)}{\lambda^4} + \frac{2 x_3(x_2 - x_3 + x_1)(x_2 - x_3 - x_1)}{x_1 \lambda^4} \right) \log(x_3)
\nonumber\\&&~~ ~~ ~~ ~~ ~~ ~~ ~~ ~~ ~~ ~~ ~~  + \frac{2 x_3 \Big( \lambda^2 - 3(x_1 - x_3)^2 + 3x_2^2\Big)}{\lvert \lambda^2 \rvert^{5/2}} \mathrm{\Psi}(x_1, x_2, x_3),\label{daoshu3}
\end{eqnarray}
\subsection{The effective Lagrangian for muon MDM}
Using the effective Lagrangian method, we adopt the dimension 6 operators
for the process $l^I\rightarrow l^I+\gamma$,
\begin{eqnarray}
&&\mathcal{O}_1^{L,R}=\frac{1}{(4\pi)^2}\bar{l}(i\mathcal{D}\!\!\!\slash)^3P_{L,R}l,~~~~~~~~~~~~~~
\mathcal{O}_2^{L,R}=\frac{eQ_f}{(4\pi)^2}\overline{(i\mathcal{D}_{\mu}l)}\gamma^{\mu}
F\cdot\sigma P_{L,R}l,
\nonumber\\
&&\mathcal{O}_3^{L,R}=\frac{eQ_f}{(4\pi)^2}\bar{l}F\cdot\sigma\gamma^{\mu}
P_{L,R} (i\mathcal{D}_{\mu}l),~~~~\mathcal{O}_4^{L,R}=\frac{eQ_f}{(4\pi)^2}\bar{l}(\partial^{\mu}F_{\mu\nu})\gamma^{\nu}
P_{L,R}l,\nonumber\\&&
\mathcal{O}_5^{L,R}=\frac{m_l}{(4\pi)^2}\bar{l}(i\mathcal{D}\!\!\!\slash)^2P_{L,R}l,
~~~~~~~~~~~~~~\mathcal{O}_6^{L,R}=\frac{eQ_fm_l}{(4\pi)^2}\bar{l}F\cdot\sigma
P_{L,R}l.
\end{eqnarray}
Here $\mathcal{D}_{\mu}=\partial_{\mu}+ieA_{\mu}$ and $P_{L,R}=\frac{1\mp\gamma_5}{2}$.
$F_{{\mu\nu}}$ represents the electromagnetic field strength. The operators $\mathcal{O}_{2,3,6}^{L,R}$ have relation with lepton MDM, which is the combination of the Wilson coefficients $C^{L,R}_{2,3,6}$.
The Wilson coefficients of the operators $\mathcal{O}_{2,3,6}^{L,R}$ satisfy the following relations
\begin{eqnarray}
&&C_{2}^{L,R} = (C_{3}^{L,R})^*, ~~~~~~C_{6}^{L,R} = (C_{6}^{L,R})^* .
\end{eqnarray}
 With the on-shell conditions for the external leptons,
 transformation of the used terms in the effective Lagrangian are obtained as\cite{Feng2009npb,feng08npb}
\begin{eqnarray}
&&C_{2}^{L,R} O_{2}^{L,R} + (C_{2}^{L,R})^* (O_{2}^{L,R})^* + C_{6}^{L,R} O_{6}^{L,R} + (C_{6}^{L,R})^* (O_{6}^{L,R})^*
\nonumber\\&&\Rightarrow \Big(C_{2}^{L,R} + (C_{2}^{L,R})^* + C_{6}^{L,R}\Big) O_{6}^{L,R} +
\Big((C_{2}^{L,R})^* + C_{2}^{L,R} + (C_{6}^{L,R})^*\Big) (O_{6}^{L,R})^*
\nonumber\\&&= \frac{e Q_{l} m_{l}}{(4\pi)^{2}} \Big[ \Re\Big(C_{2}^{L,R} + (C_{2}^{L,R})^* + C_{6}^{L,R}\Big)
 \bar{l} \sigma^{\mu\nu} l\nonumber\\&&~~~ + i \Im\Big(C_{2}^{L,R} + (C_{2}^{L,R})^* + C_{6}^{L,R}\Big) \bar{l} \sigma^{\mu\nu} \gamma_{5} l \Big] F_{\mu\nu}.
\end{eqnarray}
with $\Re(\cdots)$ denoting to take the real part of a complex number,
and $\Im(\cdots)$ denoting to take the imaginary part of a complex number.
The lepton MDM reads as
\begin{eqnarray}
&&{\cal L}_{{MDM}}={e\over4m_{l}}\;a_{l}\;\bar{l}\sigma^{\mu\nu}
l\;F_{{\mu\nu}}\label{adm}.
\end{eqnarray}
We finally get
\begin{eqnarray}
&&a_{l} = \frac{4 Q_{l} m_{l}^{2}}{(4\pi)^{2}} \Re\Big(C_{2}^{L,R} + (C_{2}^{L,R})^* + C_{6}^{L,R}\Big).
\end{eqnarray}

\subsection{Example 1: Barr-Zee type diagrams with scalar sub-loop and neutral vector boson}

\begin{figure}[h]
\setlength{\unitlength}{1mm}
\centering
\includegraphics[width=1.8in]{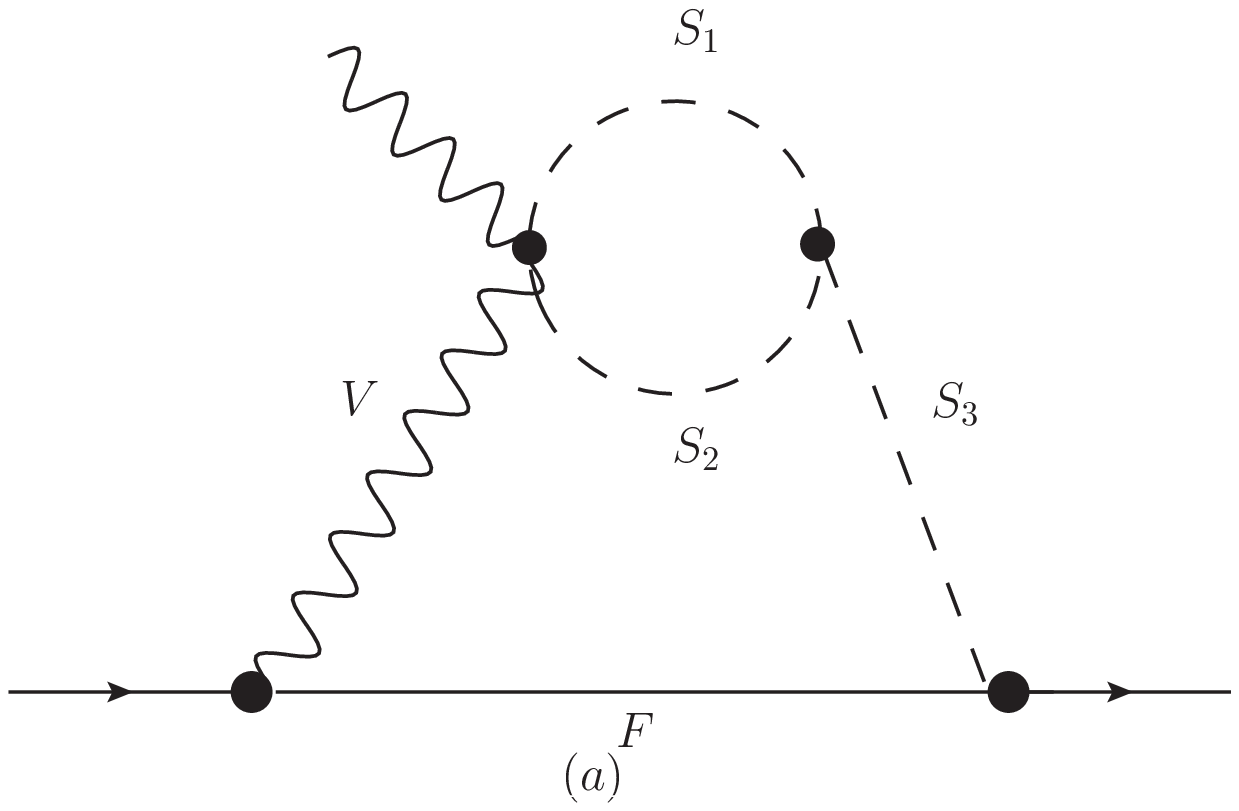}\includegraphics[width=1.8in]{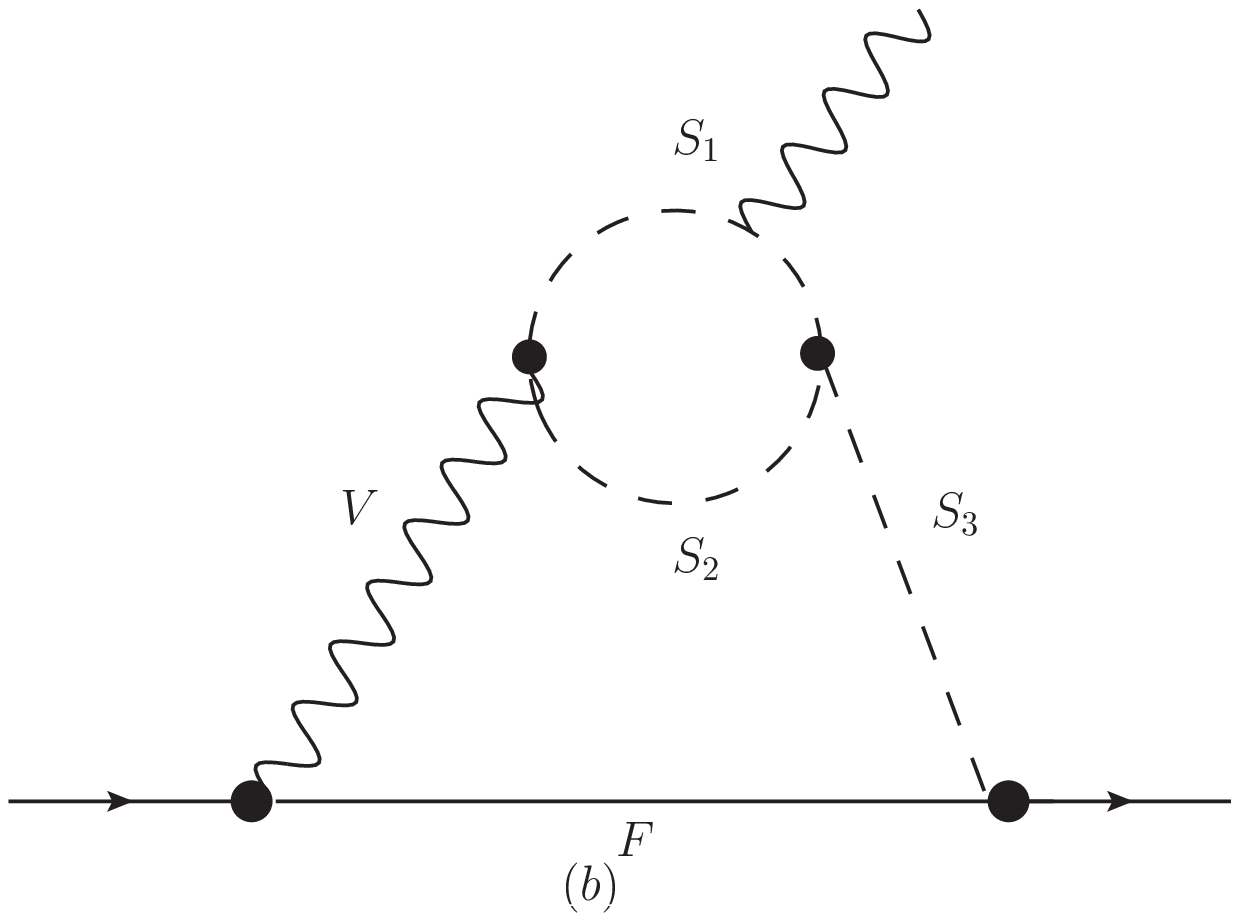}\includegraphics[width=1.8in]{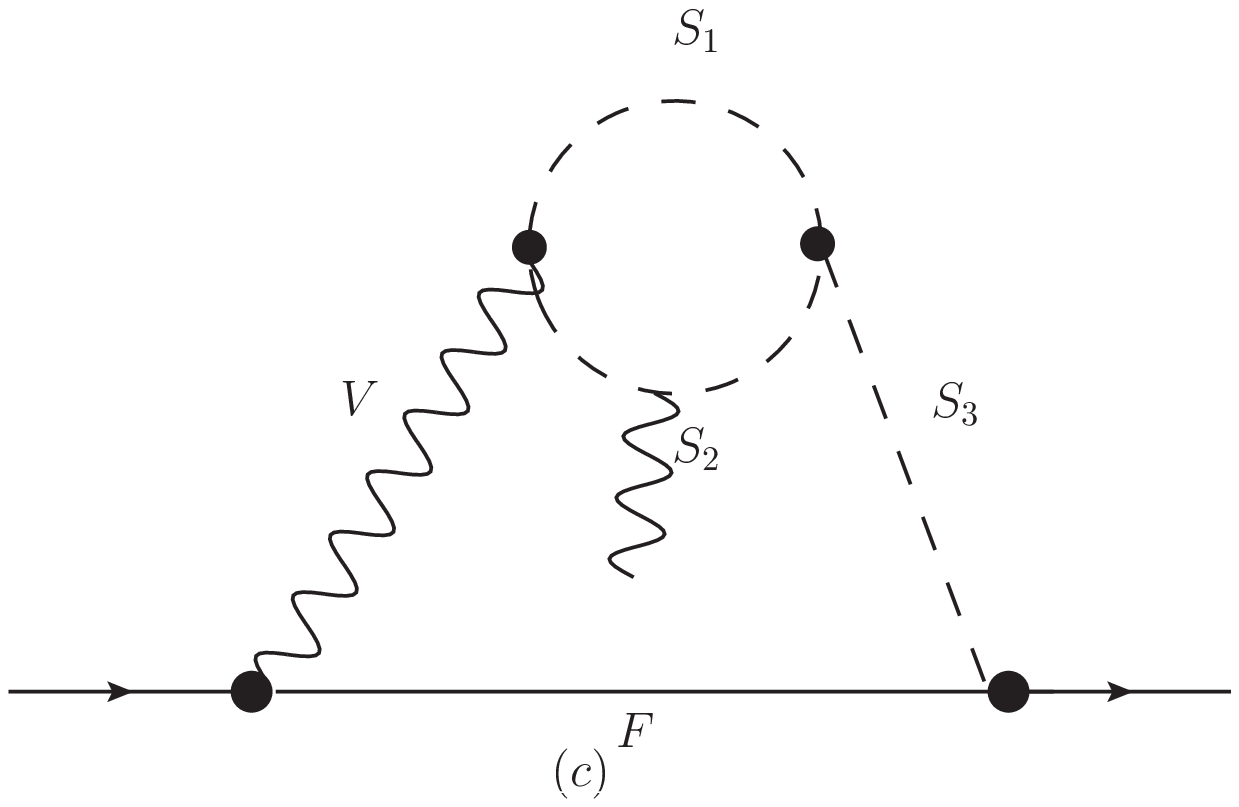}
\vspace{0cm}
\caption[]{The two loop triangle diagrams for muon MDM with neutral $V$ and $S_3$.
$S_1$ and $S_2$ in the sub-loop are charged scalar particles.}\label{BZS0}
\end{figure}

\begin{figure}[h]
\setlength{\unitlength}{1mm}
\centering
\includegraphics[width=1.8in]{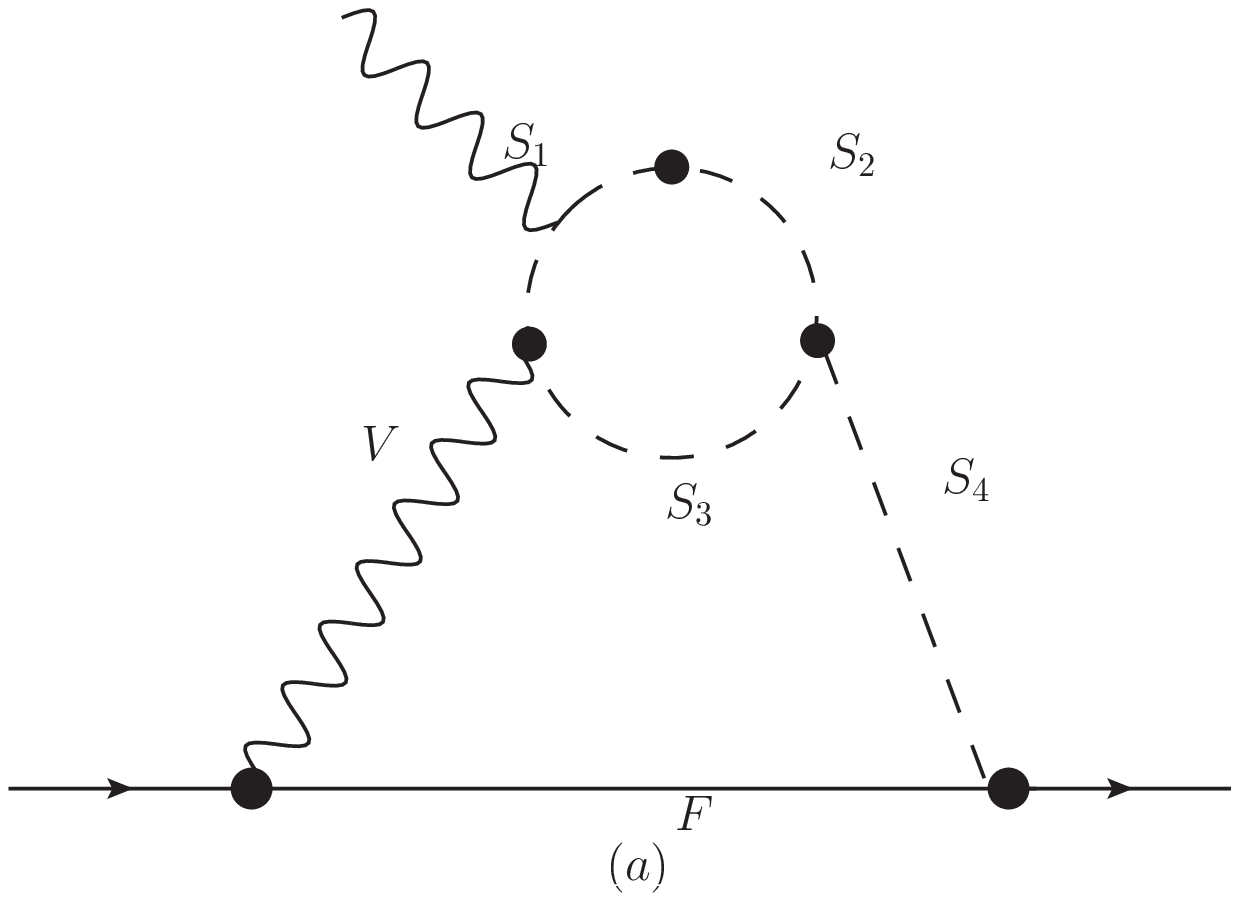},\includegraphics[width=1.8in]{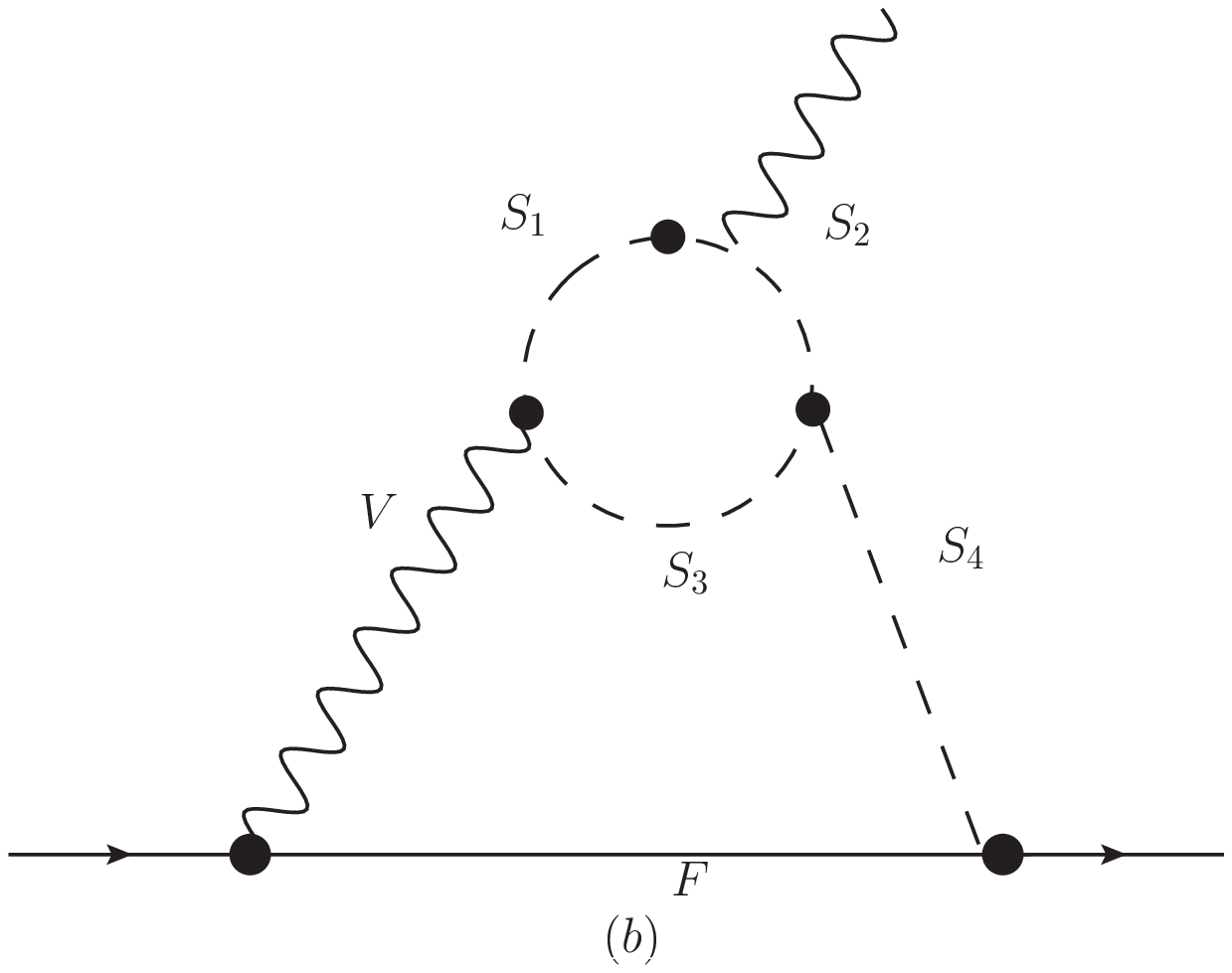}\\
\vspace{0.5cm}
\includegraphics[width=1.8in]{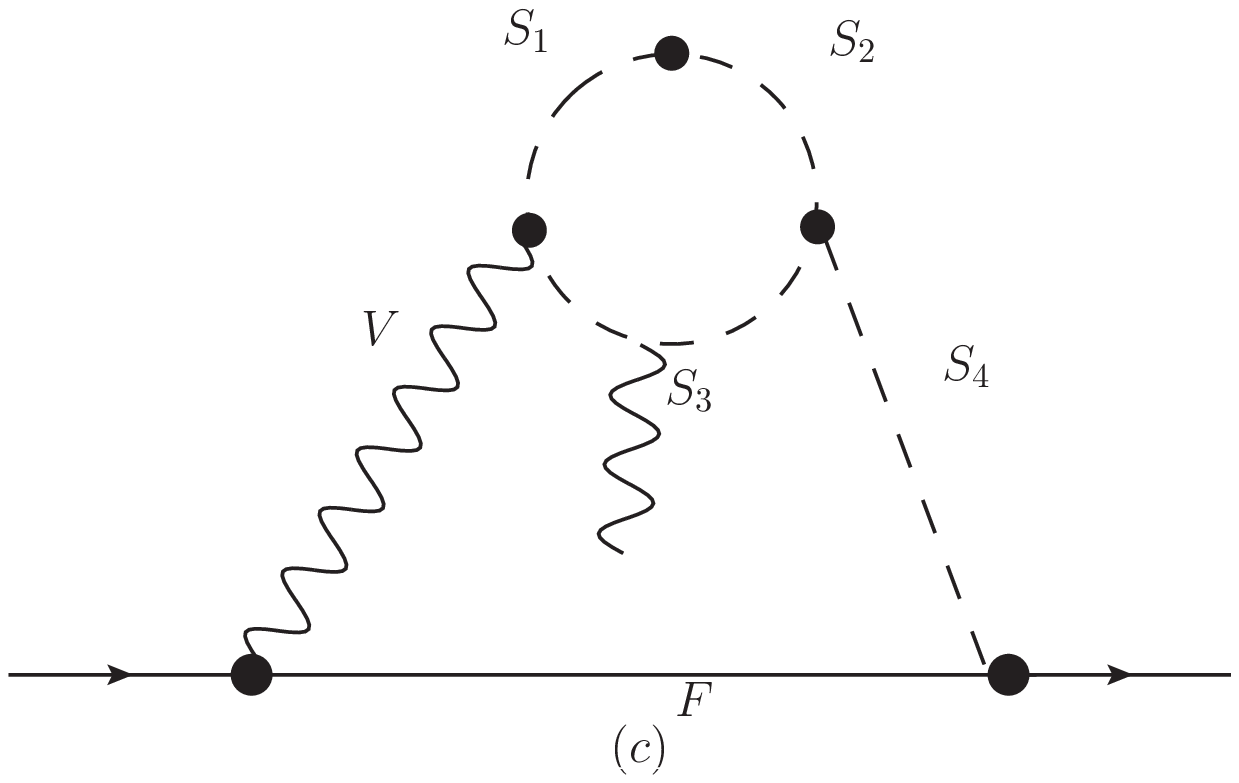},\includegraphics[width=1.8in]{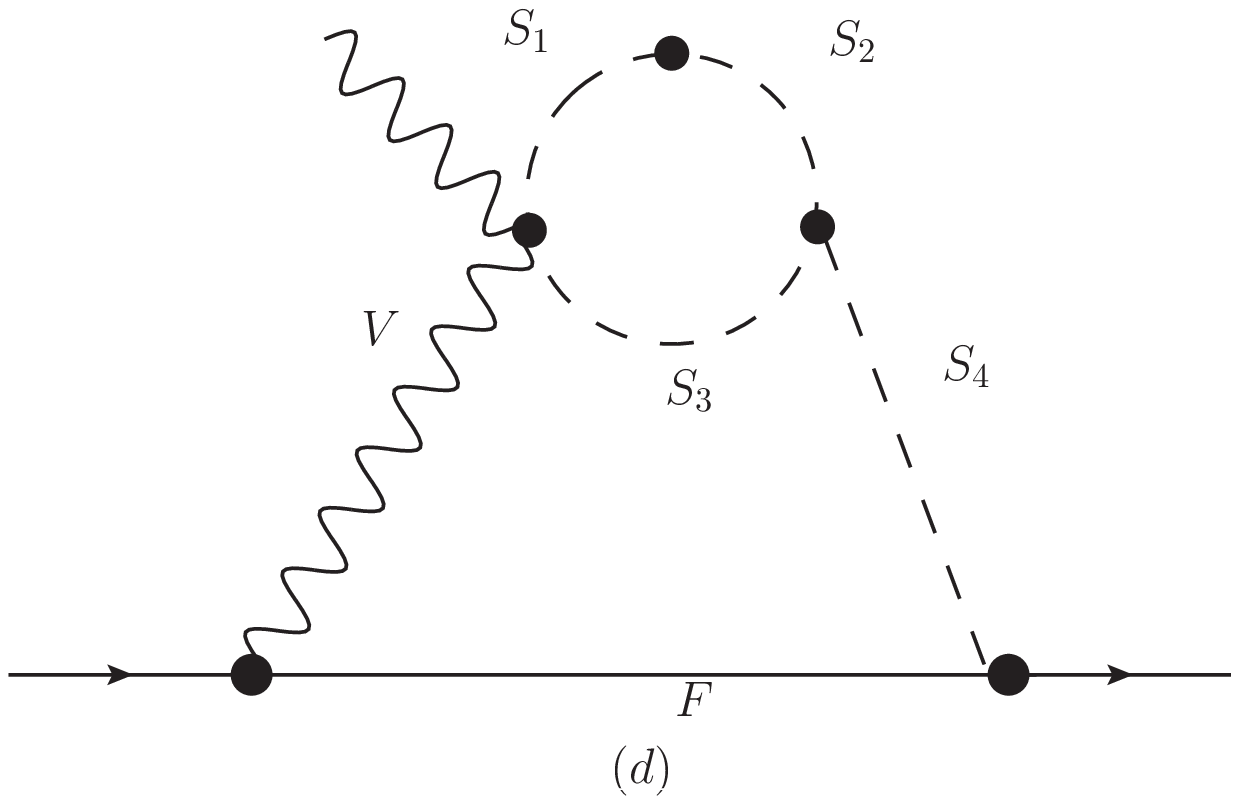}
\vspace{0cm}
\caption[]{The two loop triangle diagrams in MIA.}\label{MIASJ}
\end{figure}

\begin{figure}[h]
\setlength{\unitlength}{1mm}
\centering
\includegraphics[width=4in]{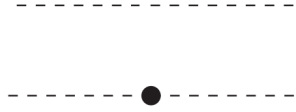}
\vspace{-11.5cm}
\caption[]{The scalar propagators appearing in MIA.}\label{chbzt}
\end{figure}

For example, we analyze Barr-Zee type diagrams with charged scalar sub-loop, neutral vector boson
and neutral Higgs. For simplicity, we use the triangle diagrams in the Fig.\ref{BZS0}.
The factor order of Fig.\ref{MIASJ} in MIA can be obtained from the operation
$\Delta M^2_{XY,S}\frac{\partial }{\partial m_{S}^2}$
on the results from the Fig.\ref{BZS0} with $X=R, L$ and $Y=R, L$.
The charged scalar masses for left-handed and right-handed are treated as same in the end.
The reason is the relation between
up scalar propagator and the down propagator with a black
spot denoting one mass insertion in MIA, and they are shown in the Fig.\ref{chbzt}.
In the Fig.\ref{chbzt}, the Feynman rule for the up propagator is $\frac{i}{k^2-m^2_S}$.
While, the Feynman rule for the down propagator
is $\frac{i}{k^2-m^2_{S_1}}\Delta M^2_{XY,S}\frac{i}{k^2-m^2_{S_2}}\sim
 \frac{i}{k^2-m^2_{S}}\Delta M^2_{XY,S}\frac{i}{k^2-m^2_{S}}$, because $m_{S_2}\sim m_{S_1}\rightarrow m_{S}$.
We take the operation $\Delta M^2_{XY,S}\frac{\partial }{\partial m_{S}^2}$ on $\frac{1}{k^2-m^2_S}$
and obtain
  \begin{eqnarray}
  \Delta M^2_{XY,S}\frac{\partial }{\partial m_{S}^2}\frac{1}{k^2-m^2_S}=\frac{1}{k^2-m^2_{S}}\Delta M^2_{XY,S}\frac{1}{k^2-m^2_{S}}.
  \end{eqnarray}
Using this relation, we can simplify the order analysis.

The sum of Fig.\ref{BZS0}(a), Fig.\ref{BZS0}(b) and Fig.\ref{BZS0}(c) satisfy
the Ward-Identity. During $C_2,~C_3$ and $C_6$, $C_6$ is the simplest one and its order is representative.
 Because $C_6$ is relatively simple and representative for the order analysis,
we calculate $C_6$ for the triangle diagrams in the Fig.(\ref{BZS0}),
and obtain the corresponding order of muon MDM.
   \begin{eqnarray}
&&C_6\propto \frac{16\pi^2}{m_F}H_{V\bar{F}F}H_{VS_1S_2}H_{S_1S_2S_3}H_{S_3\bar{F}F}
\int\int\frac{dk_1^D}{(2\pi)^D}\frac{dk_2^D}{(2\pi)^D}
\frac{1}{(k_1^2-m_{S_3}^2)(k_1^2-m_{F}^2)(k_1^2-m_{V}^2)}\nonumber\\&&\times\Big[\Big((\frac{1}{D}+\frac{1}{2})\frac{k_1\cdot k_2}{(k_2^2-m_{S_1}^2)}-\frac{2}{D}\frac{k_1\cdot k_2k_2^2}{(k_2^2-m_{S_1}^2)^2}-\frac{2}{D-1}\frac{(k_1\cdot k_2)^2}{(k_2^2-m_{S_1}^2)(k_1^2-m_{S_3}^2)}\nonumber\\&&-\frac{1}{4}\frac{k_1^2}{(k_2^2-m_{S_1}^2)}
+\frac{1}{D}\frac{k_1^2}{(k_1^2-m_{S_3}^2)}+\frac{1}{D}\frac{k_1^2k_2^2}{(k_2^2-m_{S_1}^2)^2}
+\frac{2}{D(D-1)}\frac{k_1^2k_2^2}{(k_2^2-m_{S_1}^2)(k_1^2-m_{S_3}^2)}\nonumber\\&&
+\frac{1}{D}\frac{k_1^2k_1\cdot k_2}{(k_2^2-m_{S_1}^2)(k_1^2-m_{S_3}^2)}\Big)\frac{1}{(k_2^2-m_{S_1}^2)
[(k_1-k_2)^2-m_{S_2}^2]}
%%%%%%%%%%%%%%%%%%%%%%%%%%%%%%%%%%%%%%%%%%%%%%%%%%%%%%%%
\nonumber\\&&+\Big(\frac{1}{2}\frac{k_1\cdot k_2}{(k_2^2-m_{S_2}^2)}
-\frac{2}{D}\frac{k_1\cdot k_2k_2^2}{(k_2^2-m_{S_2}^2)^2}
-\frac{2}{D-1}\frac{(k_1\cdot k_2)^2}{(k_2^2-m_{S_2}^2)(k_1^2-m_{S_3}^2)}
\nonumber\\&&
-\frac{1}{4}\frac{k_1^2}{(k_2^2-m_{S_2}^2)}
+\frac{1}{D}\frac{k_1^2k_2^2}{(k_2^2-m_{S_2}^2)^2}
+\frac{2}{D(D-1)}\frac{k_1^2k_2^2}{(k_2^2-m_{S_2}^2)(k_1^2-m_{S_3}^2)}\nonumber\\&&
+\frac{1}{D}\frac{k_1^2k_1\cdot k_2}{(k_2^2-m_{S_2}^2)(k_1^2-m_{S_3}^2)}\Big)\frac{1}{(k_2^2-m_{S_2}^2)
[(k_1-k_2)^2-m_{S_1}^2]}
\Big].
\end{eqnarray}
$H_{V\bar{F}F}$($H_{VS_1S_2}$) are the couplings of vector boson and fermion (vector boson and scalar),
and they belong to electroweak couplings at the order of $e$ $(\frac{e}{\sin\theta_W\cos\theta_W})$ with $V=\gamma(Z)$.
$H_{V\bar{F}F}$ and $H_{VS_1S_2}$ do not affect the order analysis. So we neglect them in the following discussion.

To obtain the order of $C_6^{MIA}$ for the corresponding mass insertion diagram,
we use the operation on $C_6$ as $\Delta M^2_{XY,S}\frac{\partial}{\partial m_{S_1}^2}C_6$ with $X=R, L$ and $Y=R, L$.
In the end, the simplification
$m_{S_2}\sim m_{S_1}\sim M_{NP}\gg m_V ~{\rm and} ~m_F$ are taken.
With the functions in Eq.(\ref{vacuum}), $\cdots$, Eq.(\ref{daoshu3}),
we obtain the following results through tedious calculation
\begin{eqnarray}
&&
C_6^{MIA}\propto\Delta M^2_{XY,S}\frac{\partial C_6}{\partial m_{S_1}^2}
\propto\frac{16\pi^2}{m_F}\Delta M^2_{XY,S}
H_{S_1S_2S_3}H_{S_3\bar{F}F}\frac{1}{196608 \pi ^4 M_{NP}^4}(17+\cdots )
\nonumber\\&&\hspace{1.2cm}\propto\frac{1}{12288\pi ^2 m_F M_{NP}^4}\Delta M^2_{XY,S}
H_{S_1S_2S_3}H_{S_3\bar{F}F}(17+\cdots ).
\label{C6MIA}
\end{eqnarray}
The dots in Eq.(\ref{C6MIA}) represent small terms comparing with the leading term.
As $F\rightarrow\mu$, $V\rightarrow Z(\gamma)$, $S_1\rightarrow\tilde{t}_L,~ S_2\rightarrow\tilde{t}_L,~S_3\rightarrow h_d^0$,
then
\begin{eqnarray}
&&H_{S_3\bar{F}F}\rightarrow H_{h_d^0\bar{\mu}\mu}\sim\frac{m_\mu}{v\cos\beta},\nonumber\\&&
\Delta M^2_{XY,S}\rightarrow \Delta M^2_{LR,\tilde{t}}= m_t(A_t-\mu^*\cot\beta)\sim m_tA_t,\nonumber\\&&
H_{S_1S_2S_3}\rightarrow H_{\tilde{t}_L\tilde{t}_Rh_d^0}\sim \mu m_t/v.
\end{eqnarray}
Here, we suppose $A_t\sim \mu$ and $\cot\beta$ is small.

Using the above replacements, $C_6^{MIA}$ is transformed as
\begin{eqnarray}
&&C_6^{MIA}
\propto\frac{1}{12288\pi ^2 m_\mu M_{NP}^4}m_tA_t \frac{\mu m_t}{v}\frac{m_\mu}{v\cos\beta}(17+\cdots )
\nonumber\\&&\hspace{1.2cm}\propto\frac{1}{12288\pi ^2  M_{NP}^4} \frac{A_t\mu m_t^2}{v^2 \cos\beta}(17+\cdots )\nonumber
\\&&\hspace{1.2cm}\propto\frac{\alpha_e}{12288\pi \sin\theta_W^2} \Big(\frac{A_t\mu m_t^2}{m_W^2 M_{NP}^4\cos\beta}\Big)(17+\cdots ).
 \label{C6MIA1}
\end{eqnarray}

In our discussion, $\tan\beta$ is large leading to $\sin\beta\sim1$. Therefore, $\frac{1}{\cos\beta}\sim\tan\beta$.
Then from Eq.(\ref{C6MIA}), we obtain the core factor $ \frac{A_t\mu m_t^2\tan\beta}{m_W^2 M^4_{NP}}$
neglecting other terms in the big bracket, which is consistent with the result in Ref.\cite{MEtoMIA}.
In the end, the order factor of $a_\mu$ is obtained
\begin{eqnarray}
a_\mu\propto m_\mu^2 \frac{A_t\mu m_t^2\tan\beta}{m_W^2 M^4_{NP}}
\propto \frac{m_\mu^2\tan\beta}{M^2_{NP}}.
\end{eqnarray}
Here, we use the relation $A_t\sim\mu \sim M_{NP}$.
So far, we have obtained that typical factor $\frac{m_\mu^2}{M^2_{NP}}\tan\beta$ for the Barr-Zee type two loop diagrams.

\subsection{example2: Convert the Barr-Zee type two loop diagram with heavy fermion sub-loop and photon
from mass eigenstate to MIA}
For another example, the Barr-Zee type two loop diagram with heavy fermion sub-loop and photon are studied in mass eigenstate in Ref.\cite{Feng2009npb,Zhao2022jhep}.
Regardless of whether the results obtained by mass eigenstate method or MIA method, the order of magnitude is the same,
which is a general rule. So, we transform the results of
Barr-Zee type two loop diagram with heavy fermion sub-loop and photon to MIA form, then the factor of the order is obvious.
The two-loop corrections of $a_\mu^{2L,~\gamma h^0}$ is obtained in mass eigenstate\cite{Feng2009npb,Zhao2022jhep}
\begin{eqnarray}
&&a_\mu^{2L,~\gamma h^0}=\frac{e^2}{64\sqrt{2}\pi^4}H_{h^0\bar{\mu}\mu}\sum_{F_1=F_2=\chi^\pm}\frac{x_\mu^{1/2}}{ x^{1/2}_{F_1}}
\Re(H_{h^0\bar{F}_1F_2}^L)\mathcal{ F}_1\left( x_{h}, x_{F_{\alpha}}, x_{F_{\alpha}} \right),
\\
&&\mathcal{ F}_1(x, y, z) = \frac{1}{x} \Big[
  -4(2 + \ln y)(\ln x - 1)
  - \frac{\partial}{\partial z}  (1 + 2\frac{y-z}{x}) \Phi(x, y, z)
  \nonumber\\&&~~ ~~ ~~ ~~ ~~ ~~ ~~+ \frac{\partial}{\partial z} \Big( (1 + 2\frac{y-z}{x})
   \varphi_0(y, z) + 2(y - z)\varphi_1(y, z) \Big)\Big].
\end{eqnarray}
The functions $\varphi_0(y, z)$ and $\varphi_1(y, z)$ read as
\begin{eqnarray}
&&\varphi_{0}(x, y) =
\begin{cases}
(x + y) \ln x \ln y + (x - y) \Theta(x, y), & x > y \\
2x \ln^{2} x,                               & x = y \\
(x + y) \ln x \ln y + (y - x) \Theta(y, x), & x < y
\end{cases}
\nonumber\\
&&\varphi_{1}(x, y) =
\begin{cases}
-\ln x \ln y - \frac{x + y}{x - y} \Theta(x, y), & x > y \\
4 - 2 \ln x - \ln^{2} x,                          & x = y \\
 -\ln x \ln y - \frac{x + y}{y - x} \Theta(y, x), & x < y
\end{cases}
\end{eqnarray}
The function $\Theta(x, y)$ is
\begin{eqnarray}
\Theta(x, y) = \ln x \ln\frac{y}{x} - 2 \ln(x - y) \ln\frac{y}{x} - 2 \mathrm{Li}_2\left(\frac{y}{x}\right) + \frac{\pi^2}{3}.
\end{eqnarray}
with $L_{i_2}(x)$ denoting the spence function.

Using similar assumption $m_{F_1}=m_{F_2}\gg m_{h^0}$, one can simplify the two-loop Barr-Zee type
diagrams contributing to the muon MDM\cite{Zhao2022jhep}.
\begin{eqnarray}
&&a_\mu^{2L,~\gamma h^0}=\frac{e^2}{64\sqrt{2}\pi^4}H_{h^0\bar{\mu}\mu}\sum_{F_1=F_2=\chi^\pm}\frac{x_\mu^{1/2}}{ x^{1/2}_{F_1}}
\Re(H_{h^0\bar{F}_1F_2}^L)\Big[1+\ln\frac{x_{F_1}}{x_{h^0}}\Big]. \label{algh}
\end{eqnarray}

In the MSSM, the mass matrix for charginos
in the basis ($\tilde{W}^-$,$\tilde{H}_d^-$), ($\tilde{W}^+$,$\tilde{H}_u^+$) is\cite{MSSM}
\begin{eqnarray}
M_{{\chi}^\pm}=
\left({\begin{array}{*{20}{c}}
M_2 & \frac{1}{\sqrt{2}}g_2v_u \\
\frac{1}{\sqrt{2}}g_2v_d & \mu \\
\end{array}}
\right).\label{chimass}
\end{eqnarray}
Considering that $\tan\beta=\frac{v_u}{v_d}$ is large and
$\frac{1}{\sqrt{2}}g_2v_d\ll\frac{1}{\sqrt{2}}g_2v_u\ll M_2 \sim \mu$,
the mass matrix of chargino in Eq.(\ref{chimass}) is approximately
\begin{eqnarray}
M_{{\chi}^\pm}\sim
\left({\begin{array}{*{20}{c}}
M_2 & \frac{1}{\sqrt{2}}g_2v_u \\
0 & \mu \\
\end{array}}
\right),\label{chimass1}
\end{eqnarray}
which is diagonalized by $Z_+$ and $Z_-$
\begin{eqnarray}
Z_-^\dagger M_{{\chi}^\pm} Z_+^* = M_{{\chi}^\pm}^{dia}.
\end{eqnarray}
The approximate results of the rotation matrixes $Z_-$ and $Z_+$ are deduced as
\begin{eqnarray}
&&Z_-\sim
\left({\begin{array}{*{20}{c}}
1 & \frac{ -\mu g_2v_u/\sqrt{2}}{M_2^2 - \mu^2} \\
\frac{ \mu g_2v_u/\sqrt{2}}{M_2^2 - \mu^2} & 1
\end{array}}
\right),\label{ZF}
\\&&
Z_+\sim
\left({\begin{array}{*{20}{c}}
1 & -\frac{ M_2g_2v_u/\sqrt{2}}{M_2^2 - \mu^2} \\
\frac{ M_2g_2v_u/\sqrt{2}}{M_2^2 - \mu^2} & 1
\end{array}}
\right).\label{ZZ}
\end{eqnarray}

The couplings $H_{h^0\bar{\mu}\mu}$ and $H_{h^0\bar{\chi}^\pm\chi^\pm}^L$ are
in the following form
\begin{eqnarray}
&&H_{h^0\bar{\mu}\mu}= \frac{m_l}{v\cos\beta}Z^{11}_R,
\nonumber\\
&&H^L_{h^0 \chi^\pm_i \chi^\pm_i}=\frac{-e}{\sqrt{2}s_W}(Z^{11}_R Z^{2i}_- Z^{1i}_+ + Z^{21}_R Z^{1i}_- Z^{2i}_+).\label{Hdd}
\end{eqnarray}
Here, $Z_R$ is the rotation matrix to diagonalize the mass squared matrix of CP-even Higgs.

With the Eqs.(\ref{ZF}, \ref{ZZ}, \ref{Hdd}), $a_\mu^{2L,~\gamma h^0}$ in Eq.(\ref{algh}) turns to
\begin{eqnarray}
&&a_\mu^{2L,~\gamma h^0}\propto\frac{e^2}{64\sqrt{2}\pi^4}\Big(\frac{m_\mu}{v\cos\beta}Z^{11}_R\Big)
\frac{m_\mu}{ m_{\chi}^\pm}\nonumber\\&&\times
\Re[\frac{-e}{\sqrt{2}s_W} ( Z^{1k}_R Z^{21}_- Z^{11}_+ +Z^{2k}_R Z^{11}_- Z^{21}_+
+Z^{1k}_R Z^{22}_- Z^{12}_+ +Z^{2k}_R Z^{12}_- Z^{22}_+)]
\nonumber\\&&
\propto\frac{e^2}{64\sqrt{2}\pi^4}\Big(\frac{m_\mu}{v\cos\beta}Z^{11}_R\Big)
\frac{m_\mu}{ m_{\chi}^\pm}\Re\Big[\frac{-e}{\sqrt{2}s_W}
\Big(Z^{1k}_R(Z^{21}_- + Z^{12}_+)+Z^{2k}_R(Z^{21}_+ + Z^{12}_-)\Big)\Big]
\nonumber\\&&
\propto\frac{e^2}{64\sqrt{2}\pi^4}\frac{m_\mu}{v\cos\beta}Z^{11}_R
\frac{m_\mu}{ m_{\chi}^\pm}\frac{e}{\sqrt{2}s_W} \frac{g_2v\sin\beta}{\sqrt{2}(M_2+\mu)}(Z^{11}_R-Z^{21}_R)\nonumber\\&&
\propto\frac{1}{128\sqrt{2}\pi^4}\frac{e^4 }{s_W^2}\Big(\frac{m_\mu^2\tan\beta}{m_{\chi}^\pm(M_2+\mu)}\Big)
Z^{11}_R(Z^{11}_R-Z^{21}_R)\nonumber\\&&
\propto\frac{1}{256\sqrt{2}\pi^4}\frac{e^4 }{s_W^2}
Z^{11}_R(Z^{11}_R-Z^{21}_R)\times\Big(\frac{m_\mu^2\tan\beta}{M_{NP}^2}\Big).
\label{alghs}
\end{eqnarray}
Obviously, the typical order factor $\frac{m_\mu^2\tan\beta}{M_{NP}^2}$ for $a_\mu^{2L,~\gamma h^0}$ is obtained from the above deduction.

Because to calculate the two loop
diagrams is very tedious and takes a lot of time, we give the above examples and do not show the calculation of other diagrams in detail.
We extract the order factor of the two loop diagram to find the important two loop diagrams,
which can reduce workload for the two loop research.

\subsection{The order factors for the two loop diagrams}

Similarly, we obtain the following results for Barr-Zee type two loop diagram

\begin{table}[h]
    \centering
    \caption{Barr-Zee type two loop diagrams with fermion subloop.}
    \begin{tabular}{|c|c|}
        \toprule
        \hline
        Factor & Two loop diagram \\
        \hline
        \midrule
        $\frac{m_\mu^2}{M_{NP}^2}\tan\beta$ & $F^{1.(14)}_{\chi^\pm,\chi^\pm,\mu,Z,H^0}$\raisebox{-1.0ex}{,}~$F^{1.(14)}_{\chi^\pm,\chi^0,\nu,W,H^\pm}$\\
        \hline
    \end{tabular}
    \label{T1}
\end{table}
For $F^{1.(14)}_{\chi^\pm,\chi^\pm,\mu,Z,H^0}$,
the real photon can only be attached on the virtual lepton, which does
not contribute to the lepton MDM and EDM in on-shell scheme\cite{feng08prd}.

For the Barr-Zee type two loop diagram with scalar sub loop, we show it in
 the Fig.\ref{fmtMIA2} with the particles $\mathcal{V}=\gamma, ~\mathcal{F}=\mu, ~\mathcal{S}_1=\tilde{t}_R,
~ \mathcal{S}_2=\tilde{t}_L,~ \mathcal{S}_3=\tilde{t}_R, ~\mathcal{S}_4=h^0_d$. The factor analysis is\cite{MEtoMIA}
\begin{eqnarray}
&&\frac{m_\mu^2}{M_{NP}^2}\tan\beta\frac{\mu m_t}{m_W^2}\frac{m_t(A_t-\mu^*\cot\beta)}{M_{NP}^2}
\sim\frac{m_\mu^2}{M_{NP}^2}\tan\beta\frac{\mu m_t}{m_W^2}\frac{m_tA_t}{M_{NP}^2}
\nonumber\\&&\sim
\frac{m_\mu^2}{M_{NP}^2}\tan\beta\frac{m_t^2}{m_W^2}\sim4.6\times\frac{m_\mu^2}{M_{NP}^2}\tan\beta.
\end{eqnarray}
In the same way, the following results are obtained

\begin{table}[h]
    \centering
    \caption{Barr-Zee type two loop diagrams with scalar subloop.}
    \begin{tabular}{|c|c|}
         \hline
        \toprule
        Factor & Two loop diagram  \\
        \hline
        \midrule
        $4.6\times\frac{m_\mu^2}{M_{NP}^2}\tan\beta$ & $F^{1.(15)}_{\tilde{t},\tilde{b},H^0,W,\nu}$\raisebox{-1.0ex}{,}~$F^{1.(15)}_{\tilde{t},\tilde{t},H^0,Z,\mu}$\\
        \hline
    \end{tabular}
    \label{T2}
\end{table}

After comparison, the contribution factors from
$F^{1.(15)}_{\tilde{L},\tilde{L},H^0,(\gamma,Z),\mu}$, $F^{1.(15)}_{\tilde{D},\tilde{D},H^0,(\gamma,Z),\mu}$,
$F^{1.(15)}_{\tilde{\nu},\tilde{\nu},H^0,Z,\mu}$ and $F^{1.(15)}_{\tilde{L},\tilde{\nu},H^\pm,W,{\nu}}$ are not greater than that of $F^{1.(15)}_{\tilde{t},\tilde{t},H^0,(\gamma,Z),\mu}$.
That is to say, for this type two loop diagram,
the contribution from scalar down quark, slepton and sneutrino
are not larger than that from scalar top quark.

For this type diagram, when the particles in the scalar sub-loop are Higgs,
the corresponding factors are shown as follows

\begin{table}[h]
    \centering
    \caption{Barr-Zee type two loop diagrams with Higgs subloop.}
    \begin{tabular}{|c|c|}
    \hline
        \toprule
        Factor & Two loop diagram \\
        \hline
        \midrule
        $\frac{m_\mu^2}{M_{NP}^2}\tan\beta\frac{B_\mu}{M_{NP}^2}\sim \frac{m_\mu^2}{M_{NP}^2}\tan\beta$ & $F^{1.(15)}_{H^\pm,H^\pm,H^0,(\gamma,~Z),\mu}$ $\raisebox{-1.0ex}{,}~$$F^{1.(15)}_{H^0,A^0,A^0,Z,\mu}$$\raisebox{-1.0ex}{,}~$ $F^{1.(15)}_{H^\pm,(A^0,~H^0),H^\pm,W,\nu}$\\
        \hline
    \end{tabular}
    \label{T3}
\end{table}

Here $B_\mu$ is at the order of $M_{NP}^2$.
From the above analysis, one can find $\frac{m_\mu^2}{M_{NP}^2}\tan\beta$ is the typical factor.

For a concise display of results, we collect that the diagrams possess the same factor.
To simplify the discussion, we adopt the supposition $M_Z\sim M_W\sim M_V$.

1. The diagrams have the factor $\frac{m_\mu^2}{M_{NP}^2}$\cite{Zhao2020,feng08npb} are

\begin{table}[h]
    \centering
    \caption{The two loop diagrams with small factor.}
    \begin{tabular}{|c|c|}
    \hline
        \toprule
        Factor & Two loop diagram \\
        \hline
        \midrule
        $\frac{m_\mu^2}{M_{NP}^2}$ & $F^{1.(1)}_{\gamma,S,(\gamma,~Z),\mu}$ $\raisebox{-1.0ex}{,}~$$F^{1.(9)}_{\chi^\pm,\chi^\pm,\mu,\gamma,(\gamma,~Z)}$$\raisebox{-1.0ex}{,}~$ $F^{1.(10)}_{S,S,\gamma,(\gamma,~Z),\mu}$\\
        \hline
    \end{tabular}
    \label{T4}
\end{table}

with $S$ denoting the charged scalar particles ($\tilde{L},~\tilde{U},~\tilde{D},~ H^\pm$).
The factor $\frac{m_\mu^2}{M^2_{NP}}$ does not have the large improvement term $\tan\beta$,
so these diagrams can be neglected safely.

2. The factor $\frac{m_\mu^2}{M^2_V}$ is large,
which is bigger than the factor $\frac{m_\mu^2}{M^2_{NP}}\tan\beta$ with $M_{NP}\sim 1000{\rm GeV}$.
Supposing $M_V\sim 90 {\rm GeV}$ and $\tan\beta\sim50$, the ratio of the first factor to the second factor is
\begin{eqnarray}
\frac{\frac{m_\mu^2}{M^2_V}}
{\frac{m_\mu^2}{M^2_{NP}}\tan\beta}=\frac{M^2_{NP}}{M_V^2\tan\beta}=\frac{1000^2}{90^2\times50}\sim2.47.
\end{eqnarray}
So the diagrams possessing the factor $\frac{m_\mu^2}{M^2_V}$ are important, and they are collected here.

\begin{table}[h]
    \centering
    \caption{The two loop diagrams have large factor $\frac{m_\mu^2}{M^2_V}$.}
    \begin{tabular}{|c|@{}l|}
       \hline
        \toprule
        Factor & Two loop diagram \\
         \hline
        \midrule
        $\frac{m_\mu^2}{M^2_V}$ & $F^{1.(1)}_{{Z,S,Z,\mu}}$,~$F^{1.(1)}_{W,S,W,\nu}$,~
        $F^{1.(9)}_{\chi^0,\chi^\pm,\nu,W,W}$,~$F^{1.(9)}_{\chi^\pm,\chi^\pm,\mu,Z,Z}$,~
         $F^{1.(10)}_{S,S,Z,Z,\mu}$,~$F^{1.(10)}_{S1,S2,W,W,\nu}$\\
          \hline
    \end{tabular}
    \label{T5}
\end{table}

with $S=\tilde{L},~H^\pm,~ \tilde{U},~ \tilde{D},~H^0(A^0),~\tilde{\nu}$ and $(S1,~S2)=(\tilde{\nu},~\tilde{L});~
(\tilde{U},~\tilde{D});~(H^0, ~H^{\pm})$.
In on-shell scheme, the contribution from $F^{1.(9)}_{\chi^0,\chi^0,\mu,Z,Z}$
does not have large factor $\frac{m_\mu^2}{M_V^2}$, and it is classified as negligible\cite{Feng2009npb}.

3. There are many two loop diagrams including several types that have the typical factor $\frac{m_\mu^2}{M^2_{NP}}\tan\beta$.
These two loop self-energy diagrams are collected in the table \ref{T6}, and they
account for more than half of all two loop diagrams.

\begin{table}[h]
    \centering
    \caption{The two loop diagrams have the typical factor $\frac{m_\mu^2}{M_{NP}^2}\tan\beta$.}
    \begin{tabular}{|c|@{}l|}
     \hline
        \toprule
        Factor & Two loop diagram \\
         \hline
        \midrule
        $\frac{m_\mu^2}{M_{NP}^2}\tan\beta$ & $F^{1.(2)}_{{\tilde{L},S,\tilde{L},\chi^0}}$$\raisebox{-1.0ex}{,}~$
        $F^{1.(2)}_{\tilde{\nu},S,\tilde{\nu},\chi^\pm}$$\raisebox{-1.0ex}{,}~S=\tilde{\nu},~\tilde{L},~H^0,~H^\pm$,~
        $F^{1.(3)}_{\tilde{L},\tilde{L},(\gamma,Z,W),\chi^0}$$\raisebox{-1.0ex}{,}~$$F^{1.(3)}_{\tilde{\nu},\tilde{\nu},( Z,W),\chi^\pm}$\\
        \midrule
         & $F^{1.(7)}_{\chi^0,\mu,\chi^0,\tilde{L},\tilde{L}}$$\raisebox{-1.0ex}{,}~$
         $F^{1.(7)}_{\chi^\pm,\mu,\chi^\pm,\tilde{\nu},\tilde{\nu}}$$\raisebox{-1.0ex}{,}~$
         $F^{1.(7)}_{\chi^0,\nu,\chi^\pm,\tilde{L},\tilde{\nu}}$$\raisebox{-1.0ex}{,}~$ $F^{1.(8)}_{\chi^{0},\chi^{\pm},\tilde{L},\tilde{\nu},W}$$\raisebox{-1.0ex}{,}~$$F^{1.(8)}_{\chi^{\pm,\chi^{\pm},\tilde{\nu},\tilde{\nu},Z}}$$\raisebox{-1.0ex}{,}~$$F^{1.(8)}_{\chi^{0},\chi^{0},\tilde{L},\tilde{L},Z}$\\
        \midrule
         &$F^{1.(12)}_{\mu,\chi^{\pm},\chi^{\pm},\tilde{\nu}, \tilde{\nu}}$$\raisebox{-1.0ex}{,}~$$F^{1.(12)}_{\mu,\chi^0,\chi^{0},\tilde{L},\tilde{L}}$$\raisebox{-1.0ex}{,}~$$F^{1.(12)}_{\nu,\chi^{\pm},\chi^{0},\tilde{L}, \tilde{L}}$$\raisebox{-1.0ex}{,}~$$F^{1.(12)}_{\nu,\chi^0,\chi^{\pm},\tilde{\nu},\tilde{\nu}}$$\raisebox{-1.0ex}{,}~$
         $F^{1.(13)}_{\tilde{L},\tilde{L},\tilde{L},(\gamma,~ Z),\chi^0}$
         \\
        \midrule
         &
         $F^{1.(13)}_{\tilde{\nu},\tilde{\nu},\tilde{\nu},Z,\chi^\pm}$$\raisebox{-1.0ex}{,}~$$F^{1.(13)}_{\tilde{\nu},\tilde{L},\tilde{\nu},W,\chi^\pm}$$\raisebox{-1.0ex}{,}~$
         $F^{1.(13)}_{\tilde{L},\tilde{\nu},\tilde{L},W,\chi^0}$$\raisebox{-1.0ex}{,}~$
                  $F^{1.(14)}_{\chi^\pm,\chi^\pm,\mu,Z,H^0}$$\raisebox{-1.0ex}{,}~$$F^{1.(14)}_{\chi^\pm,\chi^0,\nu,W,H^\pm}$\\
        \midrule
         &$F^{1.(15)}_{\tilde{t},\tilde{b},H^0,W,\nu}$$\raisebox{-1.0ex}{,}~$$F^{1.(15)}_{\tilde{t},\tilde{t},H^0,Z,\mu}$$\raisebox{-1.0ex}{,}~$$F^{1.(15)}_{\tilde{L},\tilde{L},H^0,(\gamma,~Z),\mu}$$\raisebox{-1.0ex}{,}~$ $F^{1.(15)}_{\tilde{D},\tilde{D},H^0,(\gamma,~Z),\mu}$$\raisebox{-1.0ex}{,}~$$F^{1.(15)}_{\tilde{\nu},\tilde{\nu},H^0,Z,\mu}$\\
        \midrule
         &$F^{1.(15)}_{\tilde{L},\tilde{\nu},H^\pm,W,\nu}$$\raisebox{-1.0ex}{,}~$$F^{1.(15)}_{H^\pm,H^\pm,H^0,(\gamma,~Z),\mu}$
         $\raisebox{-1.0ex}{,}~$$F^{1.(15)}_{H^0,A^0,A^0,Z,\mu}$$\raisebox{-1.0ex}{,}~$$F^{1.(15)}_{H^\pm,(A^0,~H^0),H^\pm,W,\nu}$\\
        \midrule
         &$F^{1.(16)}_{\chi^{0},\chi^{\pm},\chi^{0},W,\tilde{L}}$$\raisebox{-1.0ex}{,}~$$F^{1.(16)}_{\chi^{0},\chi^{0},\chi^{0},Z,\tilde{L}}$$\raisebox{-1.0ex}{,}~$$F^{1.(16)}_{\chi^{\pm},\chi^{\pm},\chi^{\pm},(\gamma,~Z), \tilde{\nu}}$$\raisebox{-1.0ex}{,}~$$F^{1.(16)}_{\chi^{\pm},\chi^{0},\chi^{\pm}, W,\tilde{\nu}}$\\
        \midrule
         &$F^{1.(18)}_{\chi^\pm,\chi^\pm,\chi^\pm,H^0,\tilde{\nu}}$$\raisebox{-1.0ex}{,}~$$F^{1.(18)}_{\chi^\pm,\chi^0,\chi^\pm,H^\pm, \tilde{\nu}}$$\raisebox{-1.0ex}{,}~$$F^{1.(18)}_{\chi^0\chi^0\chi^0H^0\tilde{L}}$$\raisebox{-1.0ex}{,}~$$F^{1.(18)}_{\chi^0,\chi^\pm,\chi^0,H^\pm,\tilde{L}}$\\
        \midrule
         &$F^{1.(18)}_{\chi^0,F,\chi^0,\tilde{S},\tilde{L}}\raisebox{-1.0ex}{,}~(F,~\tilde{S})=(\nu,~\tilde{\nu}),~ (l,~\tilde{L})$,~
         $F^{1.(18)}_{\chi^\pm,F,\chi^\pm,\tilde{S},\tilde{\nu}}\raisebox{-1.0ex}{,}~(F,~\tilde{S})=(\nu,~\tilde{L}),~ (l,~\tilde{\nu})$
         \\  \hline
    \end{tabular}
    \label{T6}
\end{table}

4. The following diagrams in table \ref{T7} have the very large factor $\frac{m_\mu^2}{M^2_V}\tan\beta$.
So these diagrams are most important to study muon MDM.

\begin{table}[h]
    \centering
    \caption{The two loop diagrams have the large factor $\frac{m_\mu^2}{M_{V}^2}\tan\beta$.}
    \begin{tabular}{|c|@{}l|}
       \hline
        \toprule
        Factor & Two loop diagram \\
        \hline
        \midrule
        $\frac{m_\mu^2}{M_{V}^2}\tan\beta$ &
        $F^{1.(4)}_{\nu,\chi^0,\chi^\pm,W,\tilde{\nu}}$$\raisebox{-1.0ex}{,}~$$F^{1.(4)}_{\nu,\chi^\pm,\chi^0,W,\tilde{L}}$$\raisebox{-1.0ex}{,}~$$F^{1.(4)}_{\mu,\chi^0,\chi^0,Z, \tilde{L}}$$\raisebox{-1.0ex}{,}~$$F^{1.(4)}_{\mu,\chi^\pm,\chi^\pm,Z,\tilde{\nu}}$\\
        \midrule
         &$F^{1.(6)}_{\chi^\pm,\mu,\tilde{\nu},\tilde{\nu},Z}$$\raisebox{-1.0ex}{,}~$$F^{1.(6)}_{\chi^0,\mu,\tilde{L},\tilde{L}, Z}$$\raisebox{-1.0ex}{,}~$$F^{1.(6)}_{\chi^0,\nu,\tilde{L},\tilde{\nu},W}$$\raisebox{-1.0ex}{,}~$
         $F^{1.(6)}_{\chi^\pm,\nu,\tilde{\nu},\tilde{L},W}$\\
        \midrule
         &$F^{1.(17)}_{\nu,\chi^{\pm},\nu,\tilde{L},W}$$\raisebox{-1.0ex}{,}~$$F^{1.(17)}_{\nu,\chi^{0},\nu,\tilde{\nu}, W}$$\raisebox{-1.0ex}{,}~$$F^{1.(17)}_{\mu,\chi^{0},\mu,\tilde{L},Z}$$\raisebox{-1.0ex}{,}~$
         $F^{1.(17)}_{\mu,\chi^{\pm},\mu,\tilde{\nu},Z}$\\
         \hline
    \end{tabular}
    \label{T7}
\end{table}
5.
This type diagrams in table \ref{T8} have the vertex $S-H-S$ possessing mass dimension, which is supposed as $\lambda_{HSS}$.
Their contributions have the factor $\frac{m_\mu^2}{M_{NP}^2}\tan\beta\times
\frac{\lambda_{HSS}}{M_{NP}}$, which is not larger than $\frac{m_\mu^2}{M_{NP}^2}\tan\beta$.
\begin{table}[h]
    \centering
    \caption{The two loop diagrams have the middle factor $\frac{m_\mu^2}{M_{NP}^2}\tan\beta
\frac{\lambda_{HSS}}{M_{NP}}$.}
    \begin{tabular}{|c|@{}l|}
        \toprule
        \hline
        Factor & Two loop diagram \\
        \hline
        \midrule
        $\frac{m_\mu^2}{M_{NP}^2}\tan\beta
\frac{\lambda_{HSS}}{M_{NP}}$ &
        $F^{1.(5)}_{\chi^\pm,\chi^\pm,\tilde{\nu},\tilde{\nu}, H^0}$$\raisebox{-1.0ex}{,}~$$F^{1.(5)}_{\chi^0,\chi^0,\tilde{L},\tilde{L},H^0}$
        $\raisebox{-1.0ex}{,}~$$F^{1.(5)}_{\chi^0,\chi^\pm,\tilde{L},\tilde{\nu},H^\pm}$
        \\ \hline
    \end{tabular}
    \label{T8}
    \end{table}

6. Similar as the above condition, this type diagram has two vertexes $H-S-S$ and the couplings $\lambda_{HSS}^2$.
From analysis, their typical factor is $\frac{m_\mu^2}{M_{NP}^2}\tan\beta\times\frac{\lambda_{HSS}^2}{M_{NP}^2}\leq \frac{m_\mu^2}{M_{NP}^2}\tan\beta.$
Because $\lambda_{HSS}$ is not larger than $M_{NP}$ in general.
\begin{table}[h]
    \centering
    \caption{The two loop diagrams have the large factor $\frac{m_\mu^2}{M_{NP}^2}\tan\beta\frac{\lambda_{HSS}^2}{M_{NP}^2}$.}
    \begin{tabular}{|c|@{}l|}
        \hline
        \toprule
        Factor & Two loop diagram \\
        \hline
        \midrule
        $\frac{m_\mu^2}{M_{NP}^2}\tan\beta\frac{\lambda_{HSS}^2}{M_{NP}^2}$ &
        $F^{1.(11)}_{\tilde{\nu},H^0,\tilde{\nu},\tilde{\nu},\chi^{\pm}}$$\raisebox{-1.0ex}{,}~$$F^{1.(11)}_{\tilde{\nu},H^\pm,\tilde{L},\tilde{\nu},\chi^{\pm}}$$\raisebox{-1.0ex}{,}~$$F^{1.(11)}_{\tilde{L},H^0,\tilde{L}, \tilde{L},\chi^{0}}$
        $\raisebox{-1.0ex}{,}~$$F^{1.(11)}_{\tilde{L},H^\pm,\tilde{\nu},\tilde{L},\chi^{0}}$
        \\ \hline
    \end{tabular}
    \label{T9}
    \end{table}

7. This type diagram $F^{1.(18)}_{F1, F2, F3, S1, S2}$ has been researched
by the authors\cite{SSL}.
In our supposition, this type diagram also has the typical factor $\frac{m_\mu^2}{M^2_{NP}}\tan\beta$.
If the internal sfermions are very heavy, they can produce  non-decoupling and logarithmically enhanced contributions to muon MDM.
Supposing that the mass of heavy squark is $M_{SH}$ and $M_{SH}\gg M_{NP}$, the logarithmically
 enhanced factor $\log\frac{M^2_{SH}}{M^2_{NP}}$ appears leading to the
  factor $\frac{m_\mu^2}{M^2_{NP}}\tan\beta \log\frac{M^2_{SH}}{M^2_{NP}}$\cite{SSL}.
  They are shown in the table \ref{T10}.

\begin{table}[h]
    \centering
    \caption{The two loop diagrams have the large factor $\frac{m_\mu^2}{M^2_{NP}}\tan\beta \log\frac{M^2_{SH}}{M^2_{NP}}$.}
    \begin{tabular}{|c|@{}l|}
    \hline
        \toprule
        Factor & Two loop diagram \\
        \hline
        \midrule
        $\frac{m_\mu^2}{M^2_{NP}}\tan\beta \log\frac{M^2_{SH}}{M^2_{NP}}$ &
        $F^{1.(18)}_{\chi^0,F,\chi^0,\tilde{S},\tilde{L}}\raisebox{-1.0ex}{,}~ (F,~\tilde{S})= (u_i,~\tilde{U}) ,~(d_i,~\tilde{D})$\\
        \midrule
         &$F^{1.(18)}_{\chi^\pm, F,\chi^\pm,\tilde{S},\tilde{\nu}}\raisebox{-1.0ex}{,}~(F,~\tilde{S})=(u_i,~\tilde{D}) ,~(d_i,~\tilde{U})$\\
         \hline
    \end{tabular}
    \label{T10}
    \end{table}

8. The diagrams in table \ref{T11} all have the lepton-Higgs-lepton vertexes, which lead to additional suppression
factor $\frac{m_\mu}{Vew}\tan\beta\lesssim0.02$\cite{MSSM}.
$Vew$ denotes the vacuum expectation value of the Higgs field with $Vew\sim 250 {\rm GeV}$.  Their total factors are shown as $\Big(\frac{m_\mu^2}{M_H^2}\tan\beta,~\frac{m_\mu^2}{M^2_{NP}}\tan\beta\Big)\times \frac{m_\mu}{Vew}\tan\beta$,
which can be neglected safely.
\begin{table}[h]
    \centering
    \caption{The two loop diagrams have the small factor suppressed by $\frac{m_\mu}{Vew}$.}
    \begin{tabular}{|c|@{}l|}
    \hline
        \toprule
        Factor & Two loop diagram \\
        \hline
        \midrule
        $\Big(\frac{m_\mu^3\tan^2\beta}{M_H^2Vew},~\frac{m_\mu^3\tan^2\beta}{M^2_{NP}Vew}\Big)$ &
        $F^{1.(7)}_{\mu,\chi^\pm,\chi^\pm,H^0,\tilde{\nu}}$$\raisebox{-1.0ex}{,}~$$F^{1.(7)}_{\mu,\chi^0,\chi^0,H^0,\tilde{L}}$$\raisebox{-1.0ex}{,}~$$F^{1.(7)}_{\nu,\chi^0,\chi^\pm,H^\pm,\tilde{\nu}}$
        $\raisebox{-1.0ex}{,}~$$F^{1.(7)}_{\nu,\chi^\pm,\chi^0,H^\pm,\tilde{L}}$\\
        \hline
    \end{tabular}
    \label{T11}
    \end{table}
This type two loop diagrams can be neglected.

9. The factor $\frac{m_\mu^2}{M^2_{NP}}\tan\beta\log x_\mu$ shown in table \ref{T12} is larger than the factor $\frac{m_\mu^2}{M^2_{NP}}\tan\beta$,
because $\log \frac{m_\mu}{M_{NP}}\sim-9$ with $M_{NP}=1000$GeV.
This factor is negative, and it is enhanced by one order of magnitude by the large logarithmic function.
This condition has been discussed by the authors\cite{twoloopphoton} for the two loop diagrams in which the internal photon couples to
at least one muon. If the internal photon is deleted, these two loop diagrams turn to the
one loop SUSY diagrams of $\mu\rightarrow \mu$, where the internal particles are slepton-neutralino and sneutrino-chargino.

\begin{table}[h]
    \centering
    \caption{The two loop diagrams have the large factor $\frac{m_\mu^2}{M^2_{NP}}\tan\beta\log x_\mu$.}
    \begin{tabular}{|c|@{}l|}
    \hline
        \toprule
        Factor & Two loop diagrams \\
        \hline
        \midrule
        $\frac{m_\mu^2}{M^2_{NP}}\tan\beta\log x_\mu$ &
        $F^{1.(4)}_{\mu,\chi^\pm,\chi^\pm,\gamma,\tilde{\nu}}$ $\raisebox{-1.0ex}{,}~$ $F^{1.(6)}_{\chi^0,\mu,\tilde{L},\tilde{L},\gamma}$ $\raisebox{-1.0ex}{,}~$ $F^{1.(17)}_{\mu,\chi^{\pm},\mu,\tilde{\nu},\gamma}$
        $\raisebox{-1.0ex}{,}~$$F^{1.(17)}_{\mu,\chi^{0},\mu,\tilde{L},\gamma}$\\
        \hline
    \end{tabular}
    \label{T12}
    \end{table}

\section{discussion and conclusion}

As it is well known that the two loop diagram contributions to muon MDM are important.
However, there are so many two loop diagrams and the calculation of each
two loop diagram is much more difficult than that of one loop diagram.
So, it becomes important to identify the magnitude of the two loop contributions.
Based on the mass eigenstate, we have analyzed the factors of the two loop triangle diagrams for $\mu\rightarrow\mu \gamma$
in the work\cite{Zhao2020}.
The effects from the rotation matrixes of the mass mixing matrixes are not taken into account,
and the obtained results are not desirable. In order to solve this shortcoming,
we use mass insertion approximation here and get overall clear factors of
the two loop diagrams for $\mu\rightarrow\mu \gamma$.

 To simplify the analysis, we suppose that
 all supersymmetric particles have the same mass $M_{NP}$ in this work.
 This approximation greatly simplifies the analysis.
 In electroweak eigenstate state, there is not the rotation matrix for supersymmetric particles.
 For the two loop diagrams in the Fig.\ref{TLSEME},
  their factors of the contributions to muon MDM are categorized by size in the table \ref{zhengl}.
 \begin{table}[h]
    \centering
    \caption{The size of the factor.}
    \begin{tabular}{|c|@{}l|}
    \hline
        \toprule
        Factor size & The factors \\
        \hline
        \midrule
        small factor &
         $\frac{m_\mu^2}{M^2_{NP}}$, $\frac{m_\mu^3\tan^2\beta}{M_H^2Vew},~\frac{m_\mu^3\tan^2\beta}{M^2_{NP}Vew}$\\
        \hline
        \midrule
        considerable factor &
        $\frac{m_\mu^2\tan\beta}{M^2_{NP}},~\frac{m_\mu^2}{M_V^2}$, $\frac{m_\mu^2\lambda_{HSS}\tan\beta}{M^3_{NP}},~\frac{m_\mu^2\lambda_{HSS}^2\tan\beta}{M^4_{NP}}$\\
        \hline
        \midrule
        large factor &
        $\frac{m_\mu^2\tan\beta}{M_V^2}$, $\frac{m_\mu^2\tan\beta}{M^2_{NP}}\log\frac{m_\mu}{M_{NP}}$, $\frac{m_\mu^2}{M^2_{NP}}\tan\beta \log\frac{M^2_{SH}}{M^2_{NP}}$\\
        \hline
    \end{tabular}
    \label{zhengl}
    \end{table}

  From these factors,
  one can find that $\frac{m_\mu^2}{M^2_{NP}}$, $\frac{m_\mu^3\tan^2\beta}{M_H^2Vew},~\frac{m_\mu^3\tan^2\beta}{M^2_{NP}Vew}$ are small, and can be ignored safely.
  The considerable factors are $\frac{m_\mu^2\tan\beta}{M^2_{NP}},~\frac{m_\mu^2}{M_V^2}$, $\frac{m_\mu^2\lambda_{HSS}\tan\beta}{M^3_{NP}},~\frac{m_\mu^2\lambda_{HSS}^2\tan\beta}{M^4_{NP}}$.
  Most two loop diagrams possess the typical factor $\frac{m_\mu^2\tan\beta}{M^2_{NP}}$.
  The rest are big ones including $\frac{m_\mu^2\tan\beta}{M_V^2}$ and $\frac{m_\mu^2\tan\beta}{M^2_{NP}}\log\frac{m_\mu}{M_{NP}}$.
  The large logarithm $\log\frac{m_\mu}{M_{NP}}$ gives negative corrections.
  As discussed in Ref.\cite{SSL}, if squarks are very heavy,
 the corrections from Fig.\ref{TLSEME}(18) type diagram are non-decoupling and logarithmically enhanced by $\frac{m_\mu^2\tan\beta}{M^2_{NP}} \log\frac{M^2_{SH}}{M^2_{NP}}$. In this condition, our results are consistent with the conclusion of the authors\cite{SSL}.
 In the end, the most important two-loop diagrams are collected in the following table \ref{MostImportant}.
  Though numerical values depend on the used models, the order analysis for the two loop
 diagrams benefits to select important diagrams from all the two loop diagrams.
 Ignoring the two loop diagrams with small factor can reduce the effort of the two loop calculations.

\begin{table}[h]
    \centering
    \caption{ The most important two loop diagrams.}
    \begin{tabular}{|c|}
    \hline

         The most important two loop diagrams \\
      \hline
      $F^{1.(4)}_{\mu,\chi^\pm,\chi^\pm,\gamma,\tilde{\nu}}$ $\raisebox{-1.0ex}{,}~$ $F^{1.(6)}_{\chi^0,\mu,\tilde{L},\tilde{L},\gamma}$ $\raisebox{-1.0ex}{,}~$ $F^{1.(17)}_{\mu,\chi^{\pm},\mu,\tilde{\nu},\gamma}$
        $\raisebox{-1.0ex}{,}~$$F^{1.(17)}_{\mu,\chi^{0},\mu,\tilde{L},\gamma}$\\
      \hline
     $F^{1.(18)}_{\chi^0,F,\chi^0,\tilde{S},\tilde{L}}\raisebox{-1.0ex}{,}~ (F,~\tilde{S})= (u_i,~\tilde{U}) ,~(d_i,~\tilde{D})$\\
      \hline
       $F^{1.(18)}_{\chi^\pm, F,\chi^\pm,\tilde{S},\tilde{\nu}}\raisebox{-1.0ex}{,}~(F,~\tilde{S})=(u_i,~\tilde{D}) ,~(d_i,~\tilde{U})$\\
         \hline
    $F^{1.(4)}_{\nu,\chi^0,\chi^\pm,W,\tilde{\nu}}$$\raisebox{-1.0ex}{,}~$$F^{1.(4)}_{\nu,\chi^\pm,\chi^0,W,\tilde{L}}$$\raisebox{-1.0ex}{,}~$$F^{1.(4)}_{\mu,\chi^0,\chi^0,Z, \tilde{L}}$$\raisebox{-1.0ex}{,}~$$F^{1.(4)}_{\mu,\chi^\pm,\chi^\pm,Z,\tilde{\nu}}$\\
       \hline
         $F^{1.(6)}_{\chi^\pm,\mu,\tilde{\nu},\tilde{\nu},Z}$$\raisebox{-1.0ex}{,}~$$F^{1.(6)}_{\chi^0,\mu,\tilde{L},\tilde{L}, Z}$$\raisebox{-1.0ex}{,}~$$F^{1.(6)}_{\chi^0,\nu,\tilde{L},\tilde{\nu},W}$$\raisebox{-1.0ex}{,}~$
         $F^{1.(6)}_{\chi^\pm,\nu,\tilde{\nu},\tilde{L},W}$\\
      \hline
         $F^{1.(17)}_{\nu,\chi^{\pm},\nu,\tilde{L},W}$$\raisebox{-1.0ex}{,}~$$F^{1.(17)}_{\nu,\chi^{0},\nu,\tilde{\nu}, W}$$\raisebox{-1.0ex}{,}~$$F^{1.(17)}_{\mu,\chi^{0},\mu,\tilde{L},Z}$$\raisebox{-1.0ex}{,}~$
         $F^{1.(17)}_{\mu,\chi^{\pm},\mu,\tilde{\nu},Z}$\\
         \hline
    \end{tabular}
    \label{MostImportant}
    \end{table}

\begin{acknowledgments}
This work is supported by National Natural Science Foundation of China (NNSFC)(No.12075074),
Natural Science Foundation of Hebei Province(A2020201002, A2022201022, A2022201017, A2023201040),
Natural Science Foundation of Hebei Education Department (QN2022173),
Post-graduate's Innovation Fund Project of Hebei University (HBU2024SS042), Scholarship Council (CSC)(No.202408130113). X. Dong acknowledges support from Funda\c{c}\~{a}o para a Ci\^{e}ncia e a Tecnologia (FCT, Portugal) through the projects CFTP FCT Unit UIDB/00777/2020 and UIDP/00777/2020.
\end{acknowledgments}

 \end{document}